\documentclass[11pt]{article}
\usepackage{psfig}
\usepackage{latexsym}
\usepackage{amssymb}
\usepackage{epsfig}
\usepackage[USenglish]{babel}
\usepackage{fullpage}

\newcommand{\comment}[1]{{}}
\newcommand{\remove}[1]{}
\newcommand{\grad}{\nabla}
\newcommand{\old}[1]{}

\newcommand{\proof}[1]{{\noindent {\it Proof.} {#1} 
        \hfill$\Box$ \vspace{1ex}}}

\newtheorem{theorem}{Theorem}
\newtheorem{lemma}[theorem]{Lemma}

\title{On the Continuous Fermat-Weber Problem\thanks{An 
extended abstract version appeared in: 16th Annual ACM Symposium on 
Computational Geometry, 2000~\cite{SoCG}.}}

\author{
{S\'andor P. Fekete}\thanks{sandor.fekete@tu-bs.de;
http://www.math.tu-bs.edu/\~{ }fekete.
Abteilung f\"ur Optimierung, Tech\-ni\-sche 
Universit\"at Braunschweig, 38106 Braunschweig, Germany.
Parts of this work were done while visiting Stony Brook University,
supported by the Deutsche Forschungsgemeinschaft, FE 407/4-1.}
\and
Joseph S. B. Mitchell\thanks{
jsbm@ams.sunysb.edu; http://www.ams.sunysb.edu/\~{ }jsbm.
De\-part\-ment of A\-pplied Ma\-the\-ma\-tics and Statistics,
State U\-ni\-ver\-si\-ty of New York, Sto\-ny Brook, NY 11794-3600.
Partially supported by the National Science Foundation (CCR-9732220,
CCR-0098172) and by grants from Honda, HRL Laboratories, Metron
Aviation, NASA (NAG2-1325), Sandia National Labs, and Sun
Microsystems.}
\and
Karin Beurer\thanks{karin.beurer@sap.com;
SAP AG, 69190 Walldorf, Germany. Contributions to this work were made
the author was working under her maiden name {\bf Weinbrecht}
at the Center for Parallel Computing, 
Universit\"at zu K\"oln, 50923 K\"oln, Germany.
}
\\
}

\date{}

\begin{document}

\maketitle
\thispagestyle{empty}

\begin{abstract}
We give the first {\em exact} algorithmic study of facility location 
problems that deal with finding a median for a 
{\em continuum} of demand points.  In particular, we consider
versions of the ``continuous $k$-median (Fermat-Weber) problem'' where the
goal is to select one or more center points that minimize the average
distance to a set of points in a demand {\em region}.  In such
problems, the average is computed as an integral over the relevant
region, versus the usual discrete sum of distances.  The resulting
facility location problems are inherently geometric, requiring
analysis techniques of computational geometry.  We provide
polynomial-time algorithms for various versions of the $L_1$ 1-median
(Fermat-Weber) problem.  We also consider the multiple-center version of the
$L_1$ $k$-median problem, which we prove is NP-hard for large~$k$.
\end{abstract}

\vfill



\newpage

\section{Introduction}

\begin{quotation}
``{\em There are three important factors that determine 
the value of real estate -- location, location, and location.}''
\end{quotation}

\paragraph{The Fermat-Weber Problem.}
There has been considerable study of facility location problems in the
field of combinatorial optimization.  In general, the input to these
problems includes a weighted set $D$ of demand locations,
with weight distribution 
$\delta$ and total weight $\mu$, a set $F$ of feasible
facility locations, and a distance function $d$ that 
measures cost between
a pair of locations. In one important class of questions, the problem
is to determine one or more feasible {\em median} locations 
$c\in C \subseteq F$
in order to minimize the average cost from the demand locations, $p\in
D$, to the corresponding central points $c_p\in C$ that are nearest to $p$:
\[ \min_{C\subset F} \frac{1}{\mu} \int\limits_{p
\in D} \delta(p) d \left(p, C \right) dp, \]
where $d(p,C)=\min_{c\in C}d(p,c)$.
If there is one median point to be placed, the problem is known as the
classical {\em Fermat-Weber problem}; its history reaches back to Fermat,
who first posed it for three points, a case that was first solved
by Torricelli. (Note that this special case has another natural generalization:
in the well-known Steiner tree problem, the objective is to find a 
connected network of minimum total length connecting a given set of points.
See \cite{fm-msmm-00} for a recent study of the relation between these problems
and further discussion.)
In the context of facility location, the median problem
was discussed in Weber's 1909
book on the pure theory of location for industries~\cite{webe} (see
\cite{w-wphp-93} for a modern survey); because of this connection, we will
speak of the {\em Fermat-Weber problem} (FWP) throughout this paper.  
More generally, for a given
number $k\geq 1$ of facilities, the problem is known as the {\em
$k$-median problem}. A problem of similar type with a different 
objective function is the so-called {\em $k$-center problem}, 
where the goal is to find a set of $k$ center locations such
that the maximum distance of the demand set from the nearest center
location is minimized.

\paragraph{Geometric Facility Location.}
There is a vast literature on location theory; for a survey, see the
book of Drezner~\cite{drez2}, with its over 1200 citations that not
only include papers dealing with mathematical aspects of optimization
and algorithms, but also various applications and heuristics.  A good
overview of research with a mathematical programming perspective is
given in the book of Mirchandani and Francis~\cite{MirFran.90}.

With many practical motivations, geometric instances of facility
location problems have attracted a major portion of the research to
date.  In these instances, the sets $D$ of demand locations and $F$ of
feasible placements are modeled as points in some geometric space,
typically $\Re^2$, with distances measured according to the Euclidean
($L_2$) or Manhattan ($L_1$) metric.  In these geometric scenarios, it
is natural to consider not only finite (discrete) sets $F$ of feasible
locations, but also (continuous) sets having positive area. For the
classical Fermat-Weber problem, the set $F$ is the entire plane $\Re^2$,
while $D$ is some finite set of demand points. 

There has been considerable activity in the computational geometry
community on facility location problems that involve computing
geometric ``centers'' and medians of various types.  The problem of
determining a 1-center, i.e., a point $c$ to minimize the maximum distance
from $c$ to a discrete set $D$ of points, is the familiar minimum
enclosing disk problem, which has linear-time algorithms based, e.g.,
on the methods of Megiddo.  The geodesic 1-center of simple polygons
has an $O(n\log n)$ algorithm~\cite{psr-cgcsp-89}; in this version
of the center problem, distances are measured according to
shortest paths (geodesics) within a simple polygon.  Recent
results of Sharir et al.~\cite{c-mptcp-99,e-fcptc-97,s-nlap2-97} have
yielded nearly-linear-time algorithms for the planar {\em two}-center
problem.  The more general $p$-center problem has been studied
recently by~\cite{sw-rpppp-96}.

\paragraph{Continuous Location Problems.}
Location theory distinguishes between discrete and continuous location
theory (see \cite{hama_nicII}).  However, for median problems,
this distinction has mostly been applied to the set of 
feasible placements, distinguishing between discrete and continuous 
sets $F$.  It is remarkable that, so far, {\em continuous} 
location theory of median problems
has almost entirely treated {\em discrete} demand
sets $D$ \cite{hama_nicII,pla2}. 
We should note that there
are several studies in the literature that deal with
$k$-center problems with continuous demand, e.g, see
\cite{mt-nrcpc-83,t-svcpclpg-87}, where demand arises
from the continuous point sets along the edges in a graph. 
See \cite{sn-nphcbpmpcgcld-88} for results on the placement
of $k$ capacitated facilities serving a continuous demand
on a one-dimensional interval.
Also, $k$-center problems have been studied extensively
in a geometric setting, see 
e.g.\ \cite{asw-d2cp-98,d-rpcp-87,h-fatcd-93,hs-bphkc-85,hlc-sdase-93,
kh-aanlp-79,ks-ckcp-00,kc-ltawt-92,klc-oaarm-90,m-we1cp-83,mz-roawe-86,
s-nlap2-97,sw-rpppp-96}. 
However, designing discrete
algorithms for $k$-center problems can generally
be expected to be more immediate than
for $k$-median problems, because the set of demand points
that determine a critical center location will usually
form just a finite set of $d+1$ points in $d$-dimensional space. 

Continuous demand for $k$-median
problems is also missing from the classification in \cite{BC89}.
We contend that the practical and
geometric motivations of the problem make it very natural to consider
exact algorithms for dealing with a continuous demand distribution
for $k$-median problems: 
if a demand occurs at some position
$p\in D$, according to some given probability density $\delta(p)$,
then we may be interested in minimizing the {\em expected} distance
$\int_{p\in D} d(c,p)\delta(p) dp$ for a feasible center location
$c\in F$.

To the best of our knowledge, there are only few references that
discuss $k$-median problems with continuous demand:
See the papers \cite{p-wcpaaglp-81,z-paglp-85}
for a discussion of continuous demand that arises probabilistically
by considering a discrete demand in an unbounded
environment with a large number of demand points,
leading to a heuristic for optimal placement of many center points.
Drezner \cite{drez3} describes in Chapter 2 of his book that normally
a continuous demand is replaced by a discrete one, for which the
error is ``quite pronounced for some problems''.
(See his chapter for some discussion of the resulting error.)
Wesolowsky and Love \cite{wl-lfrda-71} 
(and also in their book \cite{lmw-flmm-88} with Morris) 
and Drezner and Wesolowsky~\cite{dw-olfrad-80}
consider the problem
of continuous demand for rectilinear distances.
Practical motivations include the modeling of postal districts
and facility design. They compute
the optimal solution for one specific example, but fail
to give a general algorithm. More recently, Carrizosa, 
Mu\~{n}oz-M\'arquez, and Puerto~\cite{cmp-lsrfrcp-98,cmp-wprd-98} 
use convexity properties for problems of this type to deal
with the error resulting from nonlinear numerical methods
for approximating solutions. It should be noted that the objective
function is no longer convex when distances are computed in the
presence of obstacles.

In this paper, we study the $k$-median problem, and its specialization
to the Fermat-Weber problem ($k=1$), in the case of continuous demand sets.
Another way to state our continuous Fermat-Weber (1-median) problem is as
follows: In a geometric domain (e.g., cluttered with obstacles),
determine the ideal ``meeting point'' $c^*$ that minimizes the average
time that it takes an individual, initially located at a random point
in $D$, to reach $c^*$.  Another application comes from the problem of
locating a fire station in order to minimize the average distance to
points in a neighborhood, where we consider the potential emergencies
(demands) to occur at points in a continuum (the region defining the
neighborhood~$D$). As we noted above,
this objective function is different from
the situation in which we want to minimize the {\em maximum} distance
instead, a problem that has been studied
extensively in the context of discrete algorithms.

\paragraph{Choice of Metric.}
Many papers on geometric location theory have dealt with continuous
sets $F$ of feasible placements, including
\cite{ane_par,batt_gh_pa,chen_han,chep,choi_sh_ki,duri_mic,kole,kn-afrmt-03,
lar_sad,wl-lfrda-71,wes_lov2,wl-namsgwp-72}.
In the majority of these papers, distances are
measured according to the $L_1$ metric. In fact, it was shown by
Bajaj~\cite{baja} that if $L_2$ distances are used, then even in the
case of only five demand locations ($|D|=5$), the problem cannot be
solved using radicals; in particular, it cannot be solved by exact
algorithmic methods that use only ruler and compass. 
(Chandrasekaran and Tamir~\cite{ct-aofwl-90} give a polynomial-time 
approximation scheme that uses the ellipsoid method.) In this paper,
we too concentrate on the problem using the $L_1$ metric.  While we
can exactly solve some very simple special cases in the $L_2$ metric,
in general the integrations that are required to solve the problem are
likely to be just as intractable as the classic Fermat-Weber problem.  

\paragraph{Summary of Results.}
In this paper, we give the first exact algorithmic results for location
problems that are continuous on both counts, in the set $D$ as well as
the set $F$. In our model $D$ and $F$ are each given by polygonal
domains.  Our goal is to compute a set of $k$ ($k\geq 1$) optimal
centers in the feasible set $F$ that minimize the average distance
from a demand point of $D$ to the nearest center point.
Our results include:

\begin{description}
\item[(1)] A linear-time ($O(n)$) algorithm for computing an optimal
solution to the 1-median (Fermat-Weber) problem when $D=F=P$, a simple polygon
having $n$ vertices, and distance is taken to be $L_1$ geodesic
distance inside~$P$.

\item[(2)] An $O(n^2)$ algorithm for computing an optimal 1-median for
the case that $D=F=P$, a polygon with holes, and distance is taken to
be (straight-line) $L_1$ distance.

\item[(3)] An $O(I+n\log n)$ algorithm (where $I=O(n^4)$ is the
complexity of a certain arrangement) for computing an optimal 1-median
for the case that $D=F=P$, a polygon with holes, and distance is taken
to be $L_1$ geodesic distance inside~$P$.

\item[(4)] A proof of NP-hardness for the $k$-median problem when the
number of centers, $k$, is part of the input, and $D=F=P$ is a polygon
with holes. This adds specific meaning to the statement by
Wesolowsky and Love \cite{wl-lfrda-71} that computing the
optimal position of several locations ``is obviously very tedious
when (the number of locations) is very large''.

\item[(5)] Generalizations of our results to the following cases:
non-uniform probability densities over the demand set~$D$;
fixed-orientation metrics (generalization of $L_1$), which can
be used to approximate the Euclidean metric; 
higher dimensions; and, $F\neq D$ for straight-line distances.
\end{description}

\begin{figure}[htb]
 \begin{center}
  \leavevmode
  \centerline{\epsfig{file=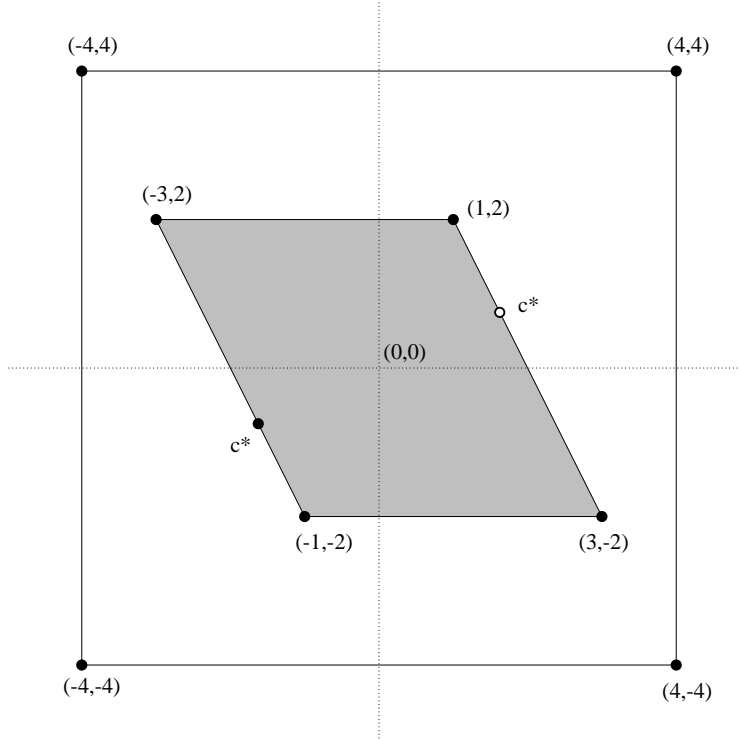, 
  width=0.60\textwidth}}
  \caption{A simple example in which the region $D=F=P$ is
a polygon (square) with a single (parallelogram) hole, shown shaded.  The average
straight-line $L_1$ distance is minimized by two optimal center points,
each marked ``$c^*$'', on the boundary of $P$.  This example is analyzed in more detail
later; refer to Figure~\ref{fig:beispiel_direkt}.)}
  \label{fig:beispiel}
 \end{center}
\end{figure}

This paper represents research done as part of the PhD thesis of
Weinbrecht; additional examples, discussion, and details may be found
in~\cite{w-ksp-99}.

\section{Preliminaries}
\label{sec:prelim}

\paragraph{Basic Definitions.}
We will let $Z=(x,y)$ denote a candidate center point in $F\subseteq
\Re^2$.  (We concentrate on two-dimensional problems until
Section~\ref{sec:conclusion}, where we discuss extensions to higher
dimensions.)  We defer discussion of multiple center points ($k>1$) to
Section~\ref{sec:many-centers}; for now, $k=1$ and we consider the
Fermat-Weber (1-median) problem.

We let $P$ denote a {\em polygonal domain}:
This is a connected planar set of points that is bounded by a finite 
set of disjoint simple closed polygonal curves. We assume that
$P$ is {\em nondegenerate}, i.e., it
is a closed set that equals the closure of its
interior points; in particular, the interior
is connected. We say that the {\em vertices} of $P$
are the vertices of its boundary; as part of the nondegeneracy assumption,
we assume that each vertex
is incident to precisely one boundary polygon and to two edges. 
In the case of one connected boundary,
we say that a polygon $P$ is {\em simple}; otherwise we say that $P$
has one or several {\em holes}, i.e, bounded components of
the complement $\Re^2\setminus P$.
A {\em critical vertex} $v$ of $P$ is one that has locally
extremal $x$- or $y$-coordinate relative to the boundary component
containing $v$, and an interior angle of at least $\pi$. 
A {\em chord} of $P$ is a straight line segment within
$P$ that connects two points on the boundary of $P$.
If $P$ is a simple polygon, any chord subdivides $P$ into two or more
pieces.

For purposes of our discussions, we focus on the case in which $D=F=P$: we
restrict $Z$ to $P$, which also equals the demand set.  
Furthermore, we focus our discussion to the case
in which the demand is uniformly distributed over the set $D=P$, so
our goal is to minimize the
average distance, $f(Z)$, given by the integral
$$f(Z)=f(x,y)=\frac{1}{\mu}{\int\!\int}_{(u,v)\in P} d(Z,(u,v)) du dv,$$
where $\mu$ is the total area of $P$,
$d(\cdot,\cdot)$ denotes either (straight-line) $L_1$ distance
or geodesic (shortest-path) 
$L_1$ distance within $P$.  (We abuse notation slightly by
writing $f(Z)=f((x,y))=f(x,y)$.)

\paragraph{Shortest Path Maps.}
In order to analyze the $k$-median problem with respect to geodesic
distances, we will utilize several definitions and results from the
theory of geometric shortest paths among obstacles; see
Mitchell~\cite{m-spn-97,m-gspno-99} for surveys on the subject of
geometric shortest paths.

\begin{figure}[htb]
 \begin{center}
  \leavevmode
  \centerline{\epsfig{file=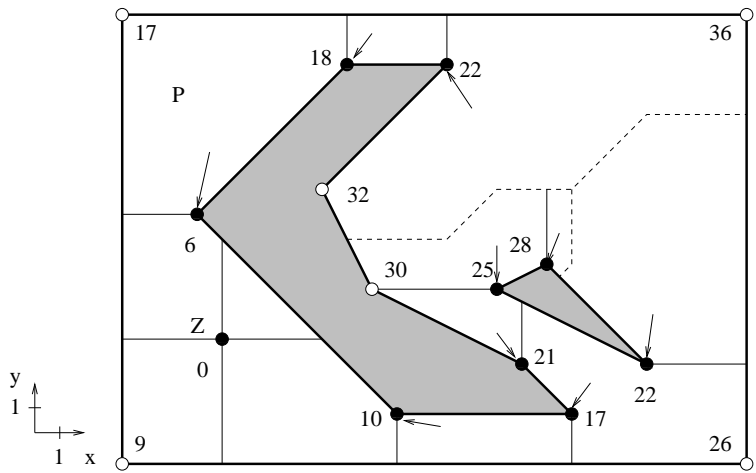, width=.6\textwidth}}
  \caption{An example of the $L_1$ geodesic SPM($Z$) 
    for a polygon $P$ with two holes (shaded).  Vertices are labeled
    with their $L_1$ geodesic distances from $Z$.  The cell rooted at
    $Z$ is partitioned into four quadrants by a vertical and a
    horizontal chord through $Z$.  Within each other cell, rooted at a
    vertex of $P$, an arrow is drawn pointing to the root of the cell.
    Vertices of $P$ whose cells are nonempty are drawn as solid black
    dots; the white vertices have empty cells in SPM($Z$).}
  \label{fig:spt_spm}
 \end{center}
\end{figure}

For a given geometric environment $P$, the {\em shortest-path map},
SPM($Z$) with respect to source point $Z$ represents the set of
shortest paths within $P$ from $Z$ to all other points $t\in P$.  The
SPM($Z$) is a decomposition of $P$ into {\em cells} $\sigma$, each
having a unique {\em root vertex}, $r_\sigma$, such that a shortest
path within $P$ from $Z$ to $t\in P$ is given by following a
shortest path within $P$ from $Z$ to $r_\sigma$, then going from
$r_\sigma$ to $t$ directly along a ``straight'' segment.  A {\em
  straight segment} between $p\in P$ and $q\in P$ is a path whose
length is $d(p,q)$, where $d(\cdot,\cdot)$ denotes our underlying
distance function ($L_1$, $L_2$, etc.). If the underlying distance is
$L_2$, a straight segment is necessarily a straight line segment; if
the underlying distance is $L_1$, as in most of this paper, a straight
segment from $p$ to $q$ consists of any path from $p$ to $q$ that is
both $x$-monotone and $y$-monotone (e.g., an ``L-shaped'' path or a
``staircase'' is straight in the $L_1$ metric).
%
%
If $t$ lies in a cell $\sigma$ rooted at $r_\sigma$, the geodesic
distance from $Z$ to $t$ is given by
$d_G(Z,t)=d_G(Z,r_\sigma)+d(r_\sigma,t)$, where $d_G(\cdot,\cdot)$
denotes the shortest-path (geodesic) distance function induced by $P$.
There is also a cell with root $Z$, consisting of points $t\in P$ for
which a shortest path from $Z$ to $t$ is attained by a straight
segment $Zt$.  Points $t$ on the common boundary of two or more cells
are optimally reached by shortest paths to $t$ whose last segment
joins $t$ to the root vertex of any one of the cells having $t$ on its
boundary. Refer to Figure~\ref{fig:spt_spm} for an example of SPM($Z$)
in the $L_1$ geodesic metric.  Each vertex $v$ of $P$ is the root of a
(possibly empty) cell.  A vertex $v=r_\sigma$ that is the root of a
cell $\sigma$ lies on the boundary of a neighboring cell $\sigma'$, so
there is a shortest path from $Z$ to $r_\sigma$ whose last segment is
$r_{\sigma'}r_\sigma$; we say that $r_{\sigma'}$ is the {\em
  predecessor} of $r_\sigma$.  In addition to its predecessor, we
store with each vertex $v$ of $P$ the length, $d_G(Z,v)$, of a
shortest path within $P$ from $Z$ to $v$.  In summary, SPM($Z$) is a
decomposition of $P$ into cells according to the ``combinatorial
structure'' (sequence of obstacle vertices along the path) of shortest
paths from $Z$ to points in the cells.

\begin{figure}[htb]
 \begin{center}
  \leavevmode
  \centerline{\epsfig{file=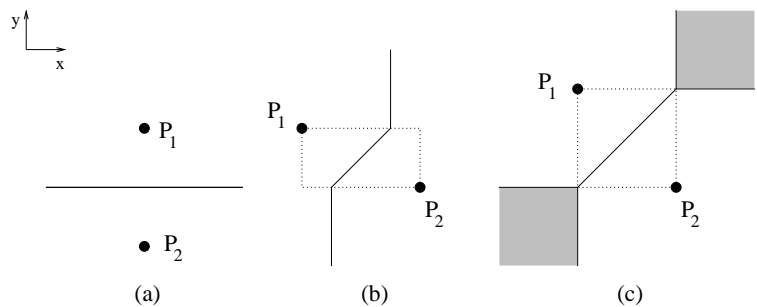,
          width=.6\textwidth}}
  \caption{Examples of the $L_1$ bisectors between two points, $P_1$ and $P_2$,
    in the $L_1$ metric.  In each of the three cases, the locus is
    shown of points $p$ for which $d(P_1,p)=d(P_2,p)$. Note that in
    case (c), when $P_1$ and $P_2$ lie on a line of slope $\pm 1$, the
    bisector includes two regions (shown shaded).}
  \label{fig:bisektoren}
 \end{center}
\end{figure}

\begin{figure}[htb]
 \begin{center}
  \leavevmode
  \centerline{\epsfig{file=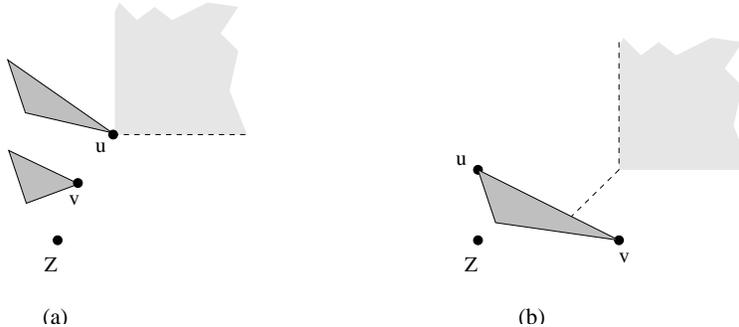,
          width=.6\textwidth}}
  \caption{Examples of $L_1$ geodesic bisectors. In both cases, the points in the
    lightly shaded quadrant are equidistant from $Z$ via $u$ and via
    $v$; our tie-breaking rule, however, assigns the shaded bisector
    points to the cell rooted at $u$, resulting in the bisector curve
    shown dashed.}
  \label{fig:spm-bisectors}
 \end{center}
\end{figure}

For any two distinct vertices $u$ and $v$ of $P$, the {\em bisector}
with respect to $u$ and $v$ is the locus of points $p\in P$ that are
(geodesically) equidistant from $Z$ via the two distinct roots, $u$
and $v$; thus, a bisector is the (possibly empty) locus of points
$p\in P$ satisfying $d_G(Z,p)=d_G(Z,u)+d(u,p)=d_G(Z,v)+d(v,p)$.  
If the underlying metric is Euclidean, bisectors are curves, easily
seen to be straight line segments or hyperbolic arcs.  In the $L_1$
geodesic metric, bisectors may be horizontal line segments, vertical
line segments, polygonal chains of segments that are horizontal,
vertical, or diagonal (with slope $\pm 1$), or they may, in a
degenerate situation, consist of {\em regions}.  See
Figure~\ref{fig:bisektoren} and Figure~\ref{fig:spm-bisectors}.  In
order that cells of the SPM($Z$) do not overlap in regions of nonzero
area, and so that the SPM($Z$) is a planar decomposition of $P$, it is
convenient to resolve degeneracies so that all bisectors are
one-dimensional (polygonal) curves, not regions.  We do this as
follows. First, if $p\in P$ satisfies
$d_G(Z,p)=d_G(Z,u)+d(u,p)=d_G(Z,v)+d(v,p)$, then we consider $p$ to be
in the cell rooted at $u$ (and not in the cell rooted at $v$) if
$d(u,p)<d(v,p)$; a consequence is that the points $p\in P$ that are in
the cell rooted at $u$ are visible to $u$ (i.e., the segment $up$ lies
in the cell). Figure~\ref{fig:spm-bisectors} shows the result of
applying this rule to two situations in which the $L_1$ geodesic
bisector would otherwise be a region.  Second, we infinitesimally
perturb the vertices of $P$ so that no two vertices lie on a common
line of slope $\pm 1$; this implies that set of points $p\in P$ for
which $d_G(Z,p)=d_G(Z,u)+d(u,p)=d_G(Z,v)+d(v,p)$ {\em and}
$d(u,p)=d(v,p)$ is a polygonal chain, not a region of nonzero area.
(All of our results apply also to the unperturbed problem instances,
by standard arguments.)

It is easily seen that a small horizontal (or vertical) shift of $Z$
by an amount $h$ results in a shift of the bisector between two
vertices $u$ and $v$ by an amount $h/2$.

Some bisector curves, such as those horizontal and vertical segments
shown solid in Figure~\ref{fig:spt_spm}, may be crossed by shortest
paths from $Z$ to points $p\in P$.  However, bisector curves that
consist of points that are the endpoints of maximal shortest
(geodesic) paths are not crossed by any shortest path; these are shown
dashed in Figure~\ref{fig:spt_spm} and are called {\em watersheds}.
(A shortest path from $Z$ to $p\in P$ is {\em maximal} if it is not
the proper subset of a shortest path from $Z$ to some other point
$p'\neq p$ of $P$.)  One can think of the watershed bisectors as
``ridges'' that partition $P$ into regions according to the
topological type (homotopy class) of shortest paths; a point on a
watershed can be reached from at least two homotopically distinct
shortest paths from $Z$.

\old{
A special type of bisector is shown in Figure~\ref{fig:spt_spm}
by the use of dashed lines. These separate adjacent cells
that are not adjacent in the shortest-path tree; thus, no shortest
path will cross them. These bisectors play the role of {\em watersheds}:
they partition the region into pieces with identical topology of their
shortest paths from $s$. They correspond to sets of holes in $P$:
for each point on a watershed, there are two topologically different
ways to reach it, meaning that the union of the paths circles
one or several holes. Again, it is easy to see that for a nondegenerate
position of $Z$, a small perturbation of $h_x$ or $h_y$ of
the $x$- or $y$-coordinate of $Z$ moves a watershed parallel to itself
by the same amount of $h_x$ or $h_y$. 
(Cf.\ Figure~\ref{fig:bisektoren}(b); see also 
Figure~\ref{fig:ableitung_geradl} (right) for an overall visualization.)
}

Shortest path maps can be computed in optimal time $O(n\log n)$ for a
polygon with $n$ vertices, both in the Euclidean metric and in the
$L_1$ metric~\cite{hs-oaesp-99,m-oasrp-89,m-lsppo-92}.  More
generally, shortest-path maps can be applied to {\em weighted region}
metrics, where the time for traveling depends on a local density
function.  For more information, see the surveys of
Mitchell~\cite{m-spn-97,m-gspno-99}.

Finally, we define one other piece of notation.  For any position $Z$
of a center, we define two subsets of the domain $P$: the set $W(Z)$
(resp., $E(Z)$) of all points $p\in P$ for which every shortest path
from $Z$ to $p$ enters the open halfplane to the left (resp., right)
of $Z$ prior to entering the open halfplane to the right (resp., left)
of $Z$.  In other words, $W(Z)$ (resp., $E(Z)$) corresponds to the set
of points $p\in P$ for which an optimal path to $p$ initially heads to
the west (resp., east).  Similarly, $P$ is subdivided into the sets
$N(Z)$ and $S(Z)$ of points for which shortest paths initially head
north versus south.  These definitions apply both to geodesic
distances, $d_G(\cdot,\cdot)$, and to straight-line distances; refer
to Figure~\ref{fig:ableitung_geradl} for illustrations.  While there
may be points $p\in P$ that are not in $W(Z)\cup E(Z)$, such points
either lie on the vertical line through $Z$ or, in the case of
geodesic distances, lie on a bisector (since such points are reached
by at least two distinct optimal paths -- one heading initially east,
one heading initially west).  By the convention and perturbation
assumption that allows us to assume that bisectors are one-dimensional
curves (rather than regions), we see that $W(Z)\cup E(Z)$ covers all
of $P$ except for a set of zero area; a similar statement holds for
$N(Z)\cup S(Z)$.  In the following, we will use $w(Z)$, $e(Z)$,
$n(Z)$, $s(Z)$ for the area of $W(Z)$, $E(Z)$, $N(Z)$, $S(Z)$,
respectively.

\old{
In particular, for any position $Z$ of a center, the region $P$ is 
subdivided into two pieces: the set $W(Z)$ 
of all points for which a shortest path from $Z$ leaves $Z$ in
a (``western'') direction with nonnegative $x$-coordinate, 
while $E(Z)$ is the set
of all points for which a shortest path from $Z$ leaves $Z$ in
an (``eastern'') direction with non-positive $x$-coordinate. 
Similarly, we define $N(Z)$ and $S(Z)$. 
(Note that this partition looks different for straight-line
and for geodesic distances; see Figure~\ref{fig:ableitung_geradl}.)
In the following, we will use $w(Z)$, $e(Z)$, $n(Z)$, $s(Z)$
for the area of $W(Z)$, $E(Z)$, $N(Z)$, $S(Z)$, respectively.
}

\section{Local Optimality Conditions}

For any given center location $Z$, the objective function value $f(Z)$
that gives the average distance from $Z$ to points of $P$ can be
evaluated by decomposing $P$ into a set of ``elementary'' pieces,
computing the average distance for each piece, and then obtaining the
total average distance as a weighted sum of the average distances for
the pieces.  In the case of straight-line $L_1$ distance, we simply
use a trapezoidization (or triangulation) of $P$ to determine the
pieces; this can be done in linear time if $P$ is simple, or in
$O(n\log n)$ time if $P$ has holes.
In the case of geodesic distance, the shortest-path map, SPM($Z$),
gives a decomposition of $P$ into cells (each of which can be refined
into triangles or trapezoids to yield a decomposition into $O(1)$-size pieces), each having
a corresponding root vertex on its boundary.  By computing the average
distance from points of a cell to the cell's root $r$ and 
adding this average to the distance $d_G(Z,r)$, and then summing over all
cells, we obtain the average geodesic distance, $f(Z)$.

The average distance associated with a single elementary piece is given by
the following result, which can be verified easily by straightforward
integration. As any region can be subdivided into a limited
number of triangles of this type, it can be used as a stepping stone for
computing the objective value for more complicated regions.

  \begin{figure}[htb]
    \begin{center}
      \leavevmode
      \centerline{\epsfig{file=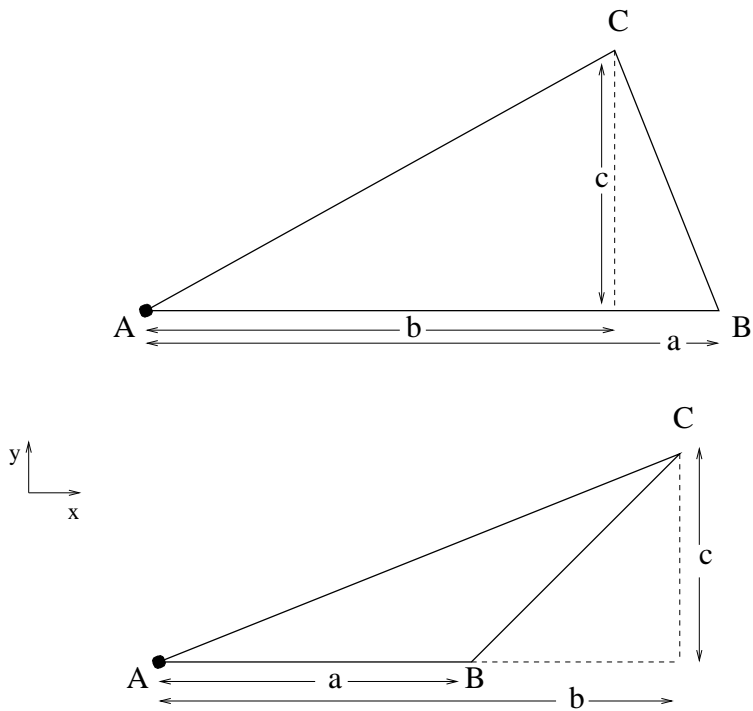,
          width=.6\textwidth}}
  \caption{Lemma~\ref{le:av.triang} notation used in computing the average $L_1$ distance of $\triangle ABC$ from the point~$A$.}
      \label{fig:abstand_allg_dreieck}
    \end{center}
  \end{figure}

\begin{lemma}
\label{le:av.triang}
For a triangle $\tau$ with vertices $A$, $B$, and $C$, such that
edge $\overline{AB}$ is horizontal,
let $a$ be the length of $\overline{AB}$,
let $c$ be the length of the altitude from $C$, and
let $b$ be the distance from $A$ to the foot of the altitude from $C$.
Then the average $L_1$ distance of points in $\tau$ from vertex $A$ is 
$\frac{1}{3}(a+b+c)$.
\end{lemma}

The objective function, $f(Z)$, is a continuous function of the
location of the center $Z$.  Because the set $F=P$ of feasible placements
is a compact domain, it follows that there is an optimum.
If $Z$ is a point minimizing $f(Z)$, then, $Z$ must be locally
optimal, meaning that there cannot be a 
feasible direction $H=(x_H,y_H)$, i.\,e., $Z+\varepsilon H \in P$
for sufficiently small $\varepsilon$, such that 
$f(Z)>f(Z+\varepsilon H)$. If $f$ is differentiable at some point
$Z\in P$, then $\langle\grad f(Z), H\rangle \geq 0$.
In particular, for interior points that are locally optimal, the
gradient must be zero; for points in the interior of boundary edges,
the gradient must be orthogonal to the boundary.
In the following lemma we compute the gradient of~$f$:

\begin{lemma}
\label{le:grad.straight}
Consider the objective function $f$ for average straight-line 
$L_1$ distance in a region $P$ of area $\mu=\mu(P)$.
Let $Z = (x, y)$ be a point in $P$.
Then the first partial derivatives
of $f$ are well-defined and given by:
\begin{equation}
\label{eq:erste_ableitung}
\begin{array}{rcl}
f_{x}(Z) & = & \frac{1}{\mu} \left( w(Z) - e(Z) \right),\\
f_{y}(Z) & = & \frac{1}{\mu} \left( s(Z) - n(Z) \right).
\end{array}
\end{equation}
\end{lemma}

\begin{figure}[htb]
 \begin{center}
  \leavevmode \centerline{\hfill
  \epsfig{file=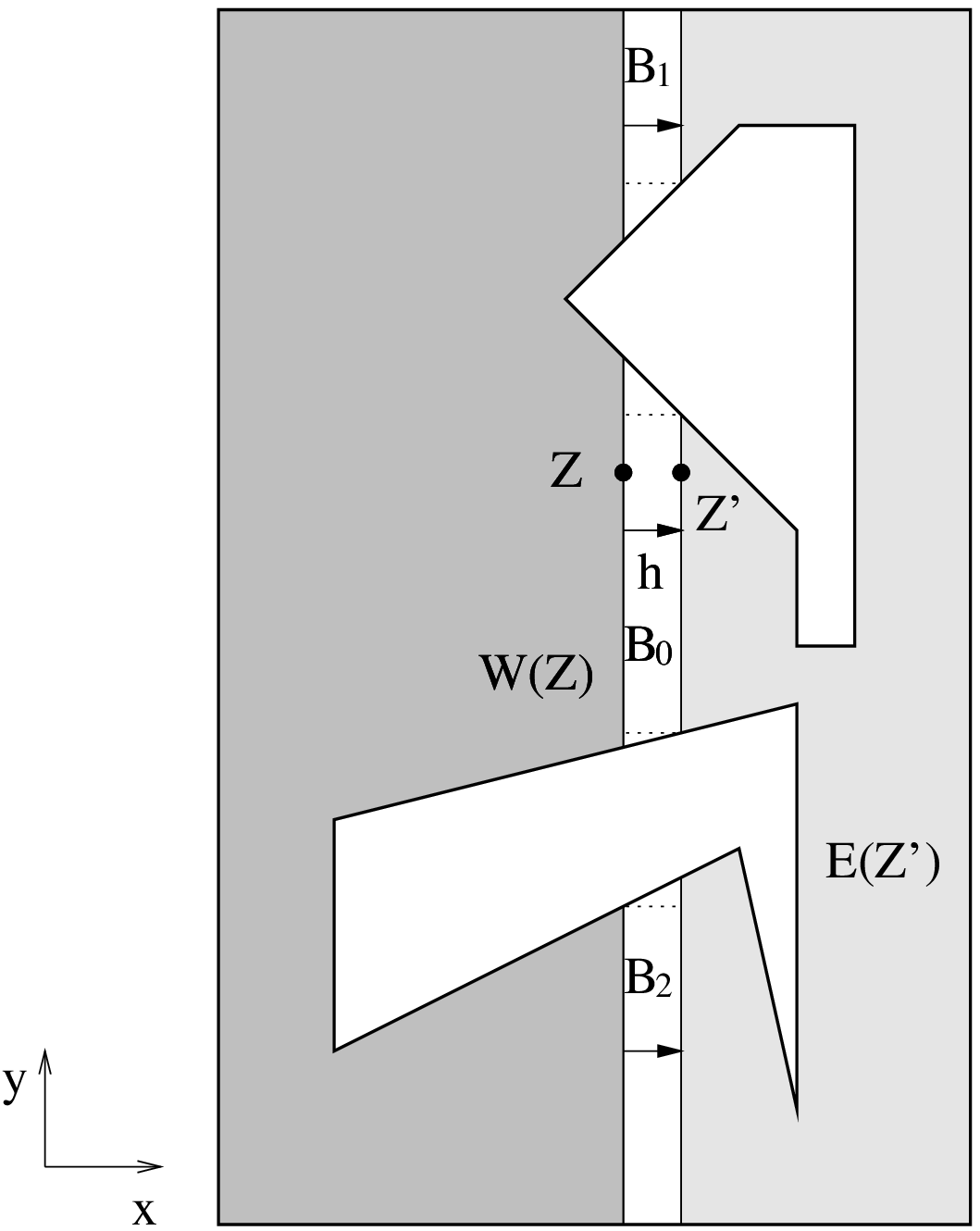, height=8cm}\hfill
  \epsfig{file=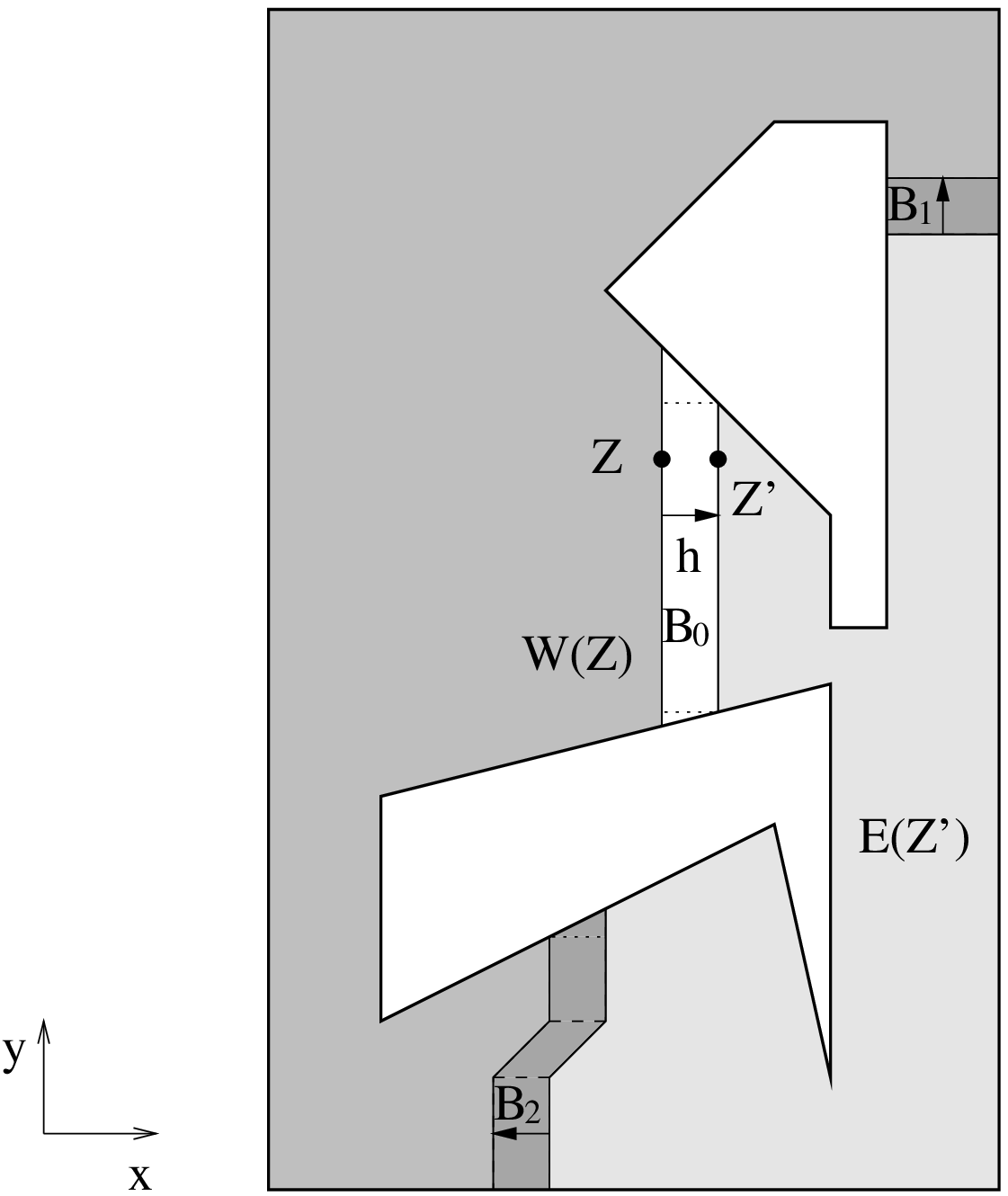, height=8cm}\hfill}
  \caption{Computing the partial derivative $f_x$ for
straight-line $L_1$ distances (on the left) and for geodesic $L_1$
  distances (on the right).}
  \label{fig:ableitung_geradl}
 \end{center}
\end{figure}

\proof{
We compute $f_x$; $f_y$ is computed similarly.
Refer to Figure~\ref{fig:ableitung_geradl} (left). 
Consider the point $Z'=Z+(h,0)$
for some sufficiently small $h$. Let $C(Z,Z'):=\{p=(x_p,y_p)\in P|
x\leq x_p\leq x+h\}$ be the narrow vertical strip between
$Z$ and $Z'$.  We compute:


\begin{eqnarray*}
&&f_{x}(x, y)  =   \lim_{h \rightarrow 0} \frac{f(x + h, y) 
- f(x, y)}{h} = \lim_{h \rightarrow 0} \frac{f(Z')-f(Z)}{h}\\
&=& \lim_{h \rightarrow 0}  \left( \frac{\frac{1}{\mu} 
\left(\int\limits_{p \in P} d \left(p,Z'\right) dp
- \int\limits_{p \in P} d \left(p,Z\right) dp\right)
}{h}\right)\\
&=& \lim_{h \rightarrow 0}  \left( \frac{\frac{1}{\mu}
\left(\int\limits_{p \in W(Z)} d \left(p,Z'\right) dp
- \int\limits_{p \in W(Z)} d \left(p,Z\right) dp\right)}{h}\right)\\
&+& \lim_{h \rightarrow 0}  \left(\frac{\frac{1}{\mu}\left(\int\limits_{p \in C(Z,Z')} d \left(p,Z'\right) dp
- \int\limits_{p \in C(Z,Z')} d \left(p,Z\right) dp\right)}{h}\right)\\
&+&\lim_{h \rightarrow 0}  \left(\frac{\frac{1}{\mu}\int\limits_{p \in E(Z')} d \left(p,Z'\right) dp
- \int\limits_{p \in E(Z')} d \left(p,Z\right) dp}{h}
\right).
\end{eqnarray*}


For small enough $h$, the area of $C(Z,Z')$ is 
a linear function of $h$, since the boundary of $P$ is
made up of straight line segments.
The term involving the difference of the two integrals over the
points $p\in C(Z,Z')$ is dependent only on the $x$-contribution
to the $L_1$ distance ($|x_p-x_Z|$ or
$|x_p-x_{Z'}|$) between $p$ and $Z$ or $Z'$, since the $y$-contributions
cancel ($y_Z=y_{Z'}$).  Since $x_p$ varies within a range of $h$,
and $|x_p-x_Z|\leq h$ (and $|x_p-x_{Z'}|\leq h$) as well, we get that
$$\left(\int\limits_{p \in C(Z,Z')} d \left(p,Z'\right) dp
- \int\limits_{p \in C(Z,Z')} d \left(p,Z\right) dp\right)\in O(h^2).$$
Since
\[\int\limits_{p \in W(Z)} d \left(p,Z'\right) dp=
\int\limits_{p \in W(Z)} \left(h+d \left(p,Z\right)\right) dp,\]
and
\[\int\limits_{p \in E(Z)} d \left(p,Z\right) dp=
\int\limits_{p \in E(Z)} \left(h+d \left(p,Z'\right)\right) dp,\]
we get
$$f_{x}(x_z, y_z)=\lim_{h \rightarrow 0}
\left(\frac{\int\limits_{p \in W(Z)}h~dp -\int\limits_{p \in E(Z)}h~dp +O(h^2)}{\mu h} \right)
= \left(\frac{1}{\mu} w(Z)-e(Z)\right),$$
as claimed.
}       

The above lemma characterizes the gradients in the case of
straight-line distances.  For geodesic distances, we proceed in a
similar manner; refer to Figure~\ref{fig:ableitung_geradl} (right).
Note that the shape of $C(Z,Z')$ is different in the case of geodesic
distances. For the example in the figure, $C(Z,Z')$ is the union of
one connected strip (shown white), bounded by vertical line segments
through $Z$ and $Z'$ and the boundary of $P$, and possibly several
narrow regions that are swept by the watersheds as $Z$ moves to $Z'$
(shown darkly shaded in the figure); note that these latter regions
only occur in the case of $P$ being a polygon with holes.

In the lemma below, we state the result under a technical assumption,
which avoids difficulties with the continuity of $w(Z)$ and $e(Z)$, as
well as with degenerate bisectors.

\begin{lemma}
\label{le:geo}
Consider the objective function $f$ for average 
geodesic $L_1$ distance in a region $P$ of area $\mu$.
Let $Z = (x_Z, y_Z)$ be a point in $P$; 
assume that neither $x_Z$ nor $y_Z$ coincide with the $x$- or $y$-coordinate 
of a critical vertex of $P$, and that $Z$ does not lie
on a watershed bisector in a shortest path map, SPM($v$), with respect
to a vertex $v$ of $P$. 
Then the first partial derivatives
of $f$ are well-defined and given by:
\begin{equation}
\label{eq:grad.geo}
\begin{array}{rcl}
f_{x}(Z) & = & \frac{1}{\mu} \left( w(Z) - e(Z) \right),\\
f_{y}(Z) & = & \frac{1}{\mu} \left( s(Z) - n(Z) \right).
\end{array}
\end{equation}
\end{lemma}

\proof{The argument is directly analogous to the proof of
Lemma~\ref{le:grad.straight}.
The assumption that $Z$ does not lie on a watershed bisector of
SPM($v$) for any vertex $v$ implies that
no watershed bisector of SPM($Z$) contains a vertex $v$ of $P$.
This implies that the
area of $C(Z,Z')$ is a linear function
of the perturbation parameter $h$. Furthermore, the difference
$d(p,Z)-d(p,Z')$ 
for points $p\in C(Z,Z')$ is bounded by $h$. The claim follows, as in the previous lemma.}

\medskip
In some situations, we make use of properties of higher-order
derivatives of $f$. In particular, we use the following
lemma:

\begin{lemma}
\label{le:cubic}
The objective function $f$ is piecewise the sum of two cubic functions,
$f_1(x)$ and $f_2(y)$, both for straight-line and for geodesic distances.
\end{lemma}


\old{
\begin{figure}[htb]
 \begin{center}
  \leavevmode
  \centerline{\epsfig{file=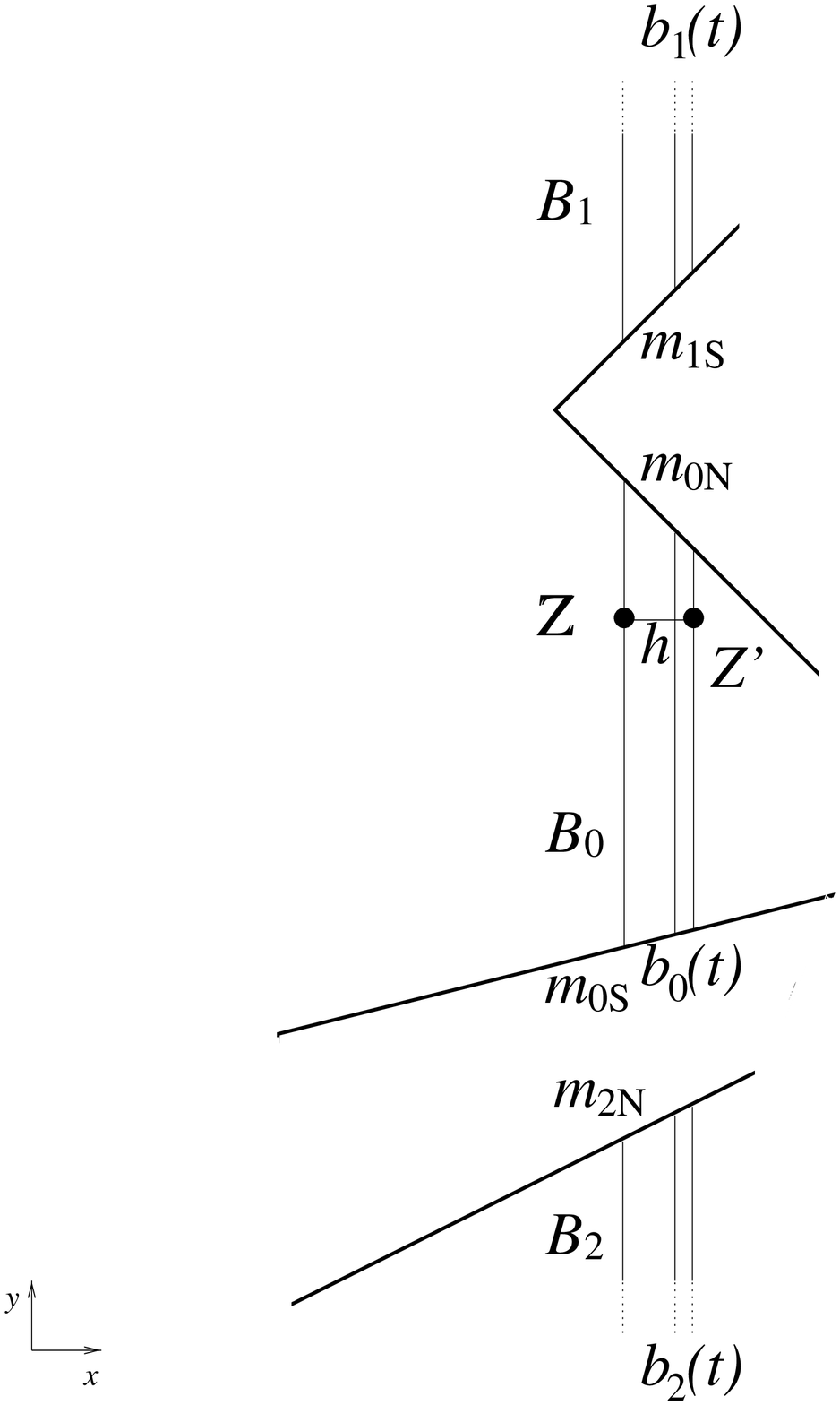,
          width=0.40\textwidth}}
  \caption{Trapezoids and slopes used in computing higher order derivatives.}
  \label{fig:ableitung_geradl_ausschnitt}
 \end{center}
\end{figure}
}

\proof{
We start by considering the objective function $f$ for average straight-line 
$L_1$ distances in $P$.
Let $Z = (x_Z, y_Z)$ be a point in $P$, with neither
$x_Z$ nor $y_Z$ coinciding with the $x$- or $y$-coordinate 
of a critical vertex of $P$.
%
%
%
Consider a small change in the $x$-coordinate of $Z$: let
$Z'=Z+(h,0)$. Then $W(Z') = W(Z) \cup C(Z,Z')$ and $E(Z') = E(Z) \cup
C(Z,Z')$, where $C(Z,Z')=B_{0} \cup B_{1} \cup \dots \cup B_{k}$ is a
union of a set of trapezoids $B_i$, each of width $h$; refer to
Figure~\ref{fig:ableitung_geradl}.  Since the area of each trapezoid
$B_i$ is a quadratic function of the width $h$, for small enough $h$,
we get that $w(Z')-w(Z)$ and $e(Z')-e(Z)$ are also quadratic in $h$.
Since $f_x(Z)=\frac{1}{\mu}(w(Z)-e(Z))$, by Lemma~\ref{le:geo}, this
implies that $f_{xxx}(Z)$ is a constant.  (Specifically,
$f_{xxx}(Z)=(2/\mu)\sum_i (m_{i,t}-m_{i,b})$, where $m_{i,t}$ (resp.,
$m_{i,b}$) is the slope of the edge of $P$ bounding the top (resp.,
bottom) of trapezoid $B_i$.) 
To see that $f$ does not contain any terms that have $xy$ as a factor, 
note that for fixed $x$, $W(Z)$ and $E(Z)$, and thus $f_x(Z)$ does not depend
on $y$. 

For geodesic distances, the claim follows in a similar manner.
Now, the regions $B_i$ forming the connected components of $C(Z,Z')$
are not necessarily vertical-walled trapezoids, but have parallel
walls formed by
translates of watershed bisectors.  See Figure~\ref{fig:ableitung_geradl}. 
Instead of being trapezoids bounded by vertical chords through
$Z$ and $Z'$, however, 
the areas $B_i$ are 
bounded by two bisectors, corresponding to translates of watershed bisectors
for $Z$ and $Z'$.
Specifically, it follows from properties
of $L_1$ bisectors described in Section~\ref{sec:prelim}
that these bisectors move in a parallel fashion,
provided that no degeneracy of a bisector occurs during the move
from $Z$ to $Z'$, i.e., no polygon vertex is hit by a bisector.
The area of region $B_i$ is thus quadratic in $h$, and the result
follows. Again, for fixed $x$, $f_x(Z)$ does not depend on $y$.
}

\section{Straight-Line $L_1$ Distance}

\begin{figure}[htb]
  \begin{center}
    \leavevmode
    \centerline{\epsfig{file=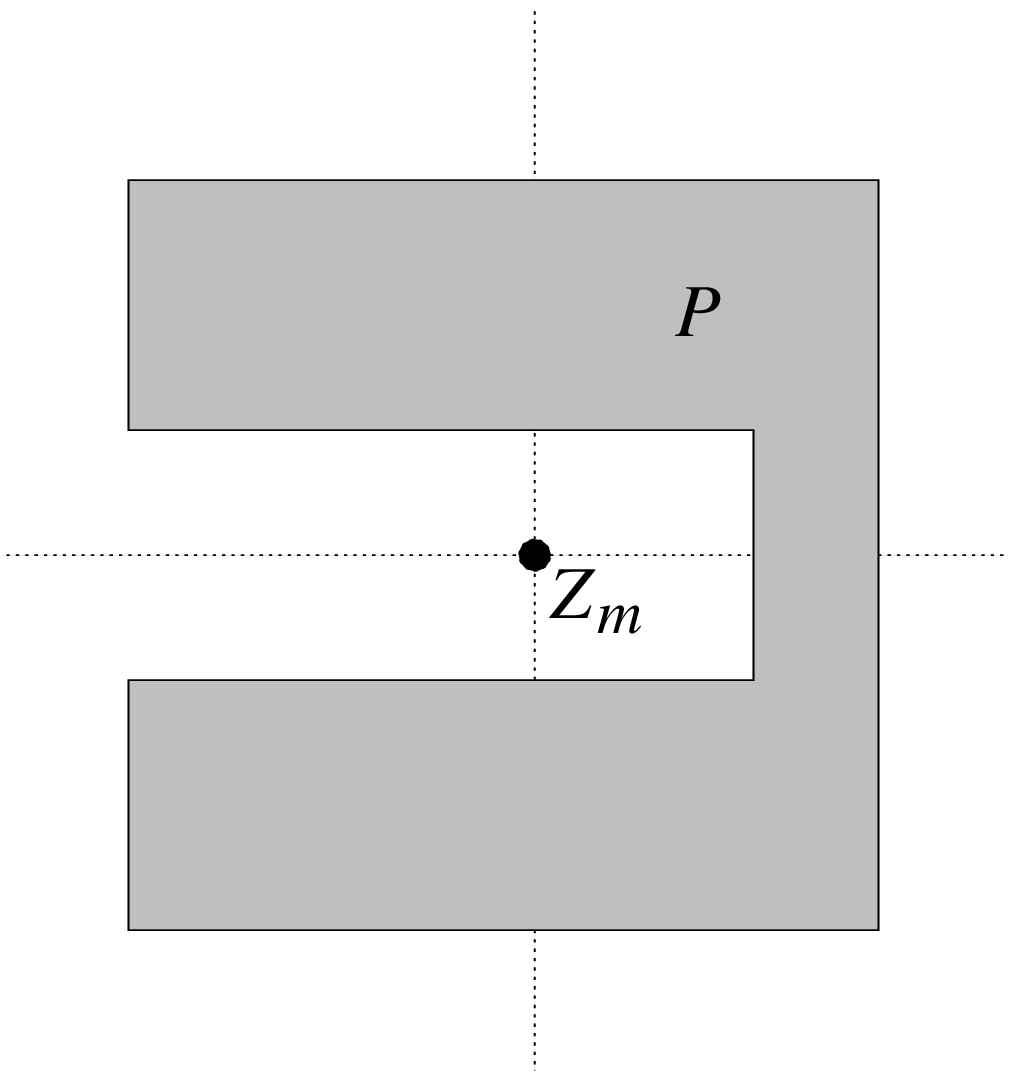,
        width=0.40\textwidth}}
    \caption{For straight-line distances, the $L_1$ origin of 
      a simple polygon $P$ may be infeasible.}
    \label{fig:straight.simple}
  \end{center}
\end{figure}

For straight-line distances, a finite average is guaranteed
even for disconnected regions $P$, as long as they are compact. 
In the following, we will consider local optimality for finding a global optimum
of $f$. Lemma~\ref{le:grad.straight} motivates considering the
$L_1$ origin of $P$, which is
a point $Z$ with $w(Z)=e(Z)$ and $n(Z)=s(Z)$, i.e., the (unique)
point that is both a median of the $x$- and the $y$-distribution.
However, the example in Figure~\ref{fig:straight.simple}
shows that even for the special case of a simple polygon $P$,
the $L_1$ origin of $P$ may not be a feasible point.

This makes it slightly involved to compute all local optima.
In the following, we describe how to evaluate them in
$O(n^2)$ time.

\begin{theorem}
\label{th:straight}
For straight-line $L_1$ distances, a point $Z^*=(x^*,y^*)$
in a polygonal region $P$ that minimizes the average 
distance $f$ to all points in $P$ can be found in
time $O(n^2)$.
\end{theorem}

\proof{
We apply the local optimality conditions. We start by
computing in time $O(n \log n)$
the $L_1$ origin $Z_m$ of $P$; if $Z_m$ is feasible (i.e., $Z_m\in P$),
we are done. 
If no interior point of $P$ is a local optimum, then we have to
consider the boundary of $P$.  This yields
a set $E$ of $O(n)$ line segments and a set $V$
of $O(n)$ vertices that we examine for local optimality.

We overlay the set of vertical and horizontal lines
through all vertices of $P$ with $E$,
subdividing each segment in $E$ into $O(n)$
pieces, bounded by a total of $O(n^2)$ ``overlay''
vertices $V_o$. Let $E_o$ be the resulting set of 
$O(n^2)$ subsegments. See Figure~\ref{fig:aufteilung}.

\begin{figure}[htb]
 \begin{center}
  \leavevmode
  \centerline{\epsfig{file=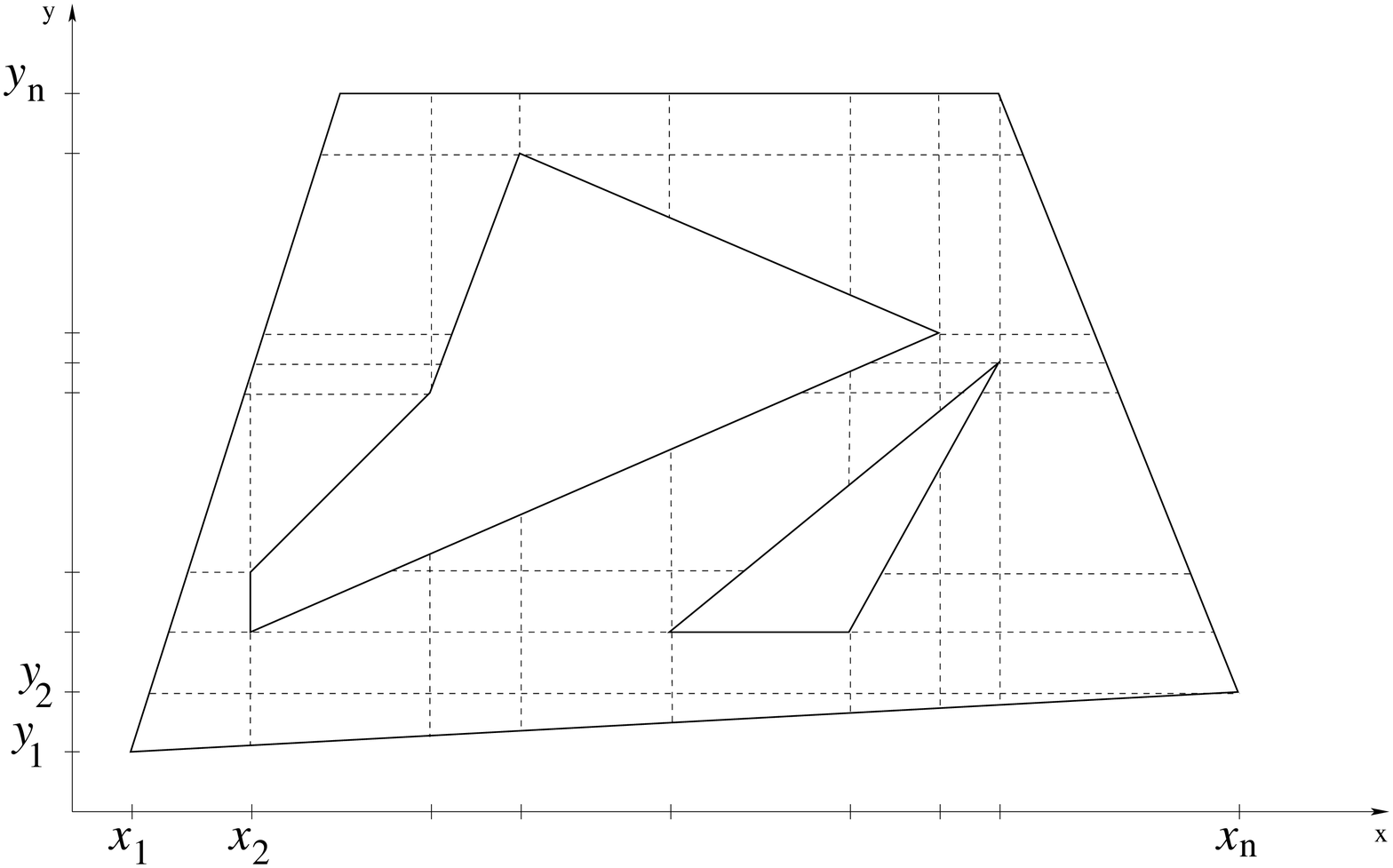, 
      width=0.60\textwidth}}
  \caption{Subdivision of the polygon into cells.}
  \label{fig:aufteilung}
 \end{center}
\end{figure}

Now we can examine the interior points $p_t=(t,y(t))$ 
of each edge $e_j\in E_o$ for local optimality. 
Let $s_j$ be a vector parallel to $e_j$.
By construction of $E_o$,
the vertical and horizontal lines through $p_t$
cannot encounter a vertex of $P$ as $p_t$ slides along one subsegment of $E_o$.
Thus, it follows from Lemma~\ref{le:cubic}
that $f_x(p_t)$ is a quadratic function in $t$,
and so is $f_y(p_t)$. Therefore, considering for $Z\in e_j$
the local optimality condition
\[\langle \grad f(p_t), s_j\rangle = 0\]
for all $O(n^2)$ subsegments $e_j\in E_o$ 
yields a set of $O(n^2)$ quadratic equations in $t$.  These can be solved
in amortized time $O(n^2)$, because we can obtain the coefficients of each quadratic equation
in amortized constant time by advancing from cell to cell in the overlay arrangement.
This gives, for each subsegment $e_j$,
at most two local optima, $q_{j, 1}$ and $q_{j, 2}$. 
Let $V_{\ell}$ be the union of $E_o$ and all $q_{j, 1}$ and $q_{j, 2}$.
By construction, $V_{\ell}$ contains $O(n^2)$
elements, and all local optima of $f$ occur at points of $V_{\ell}$.
Thus, our goal is to evaluate the objective function
at each of these points and to select the best one.  This is
simply done in amortized time $O(1)$ per candidate, by walking over
the overlay arrangement and incrementally updating the value of the
objective function.
}

\begin{figure}[htb]
  \begin{center}
    \leavevmode
    \centerline{\epsfig{file=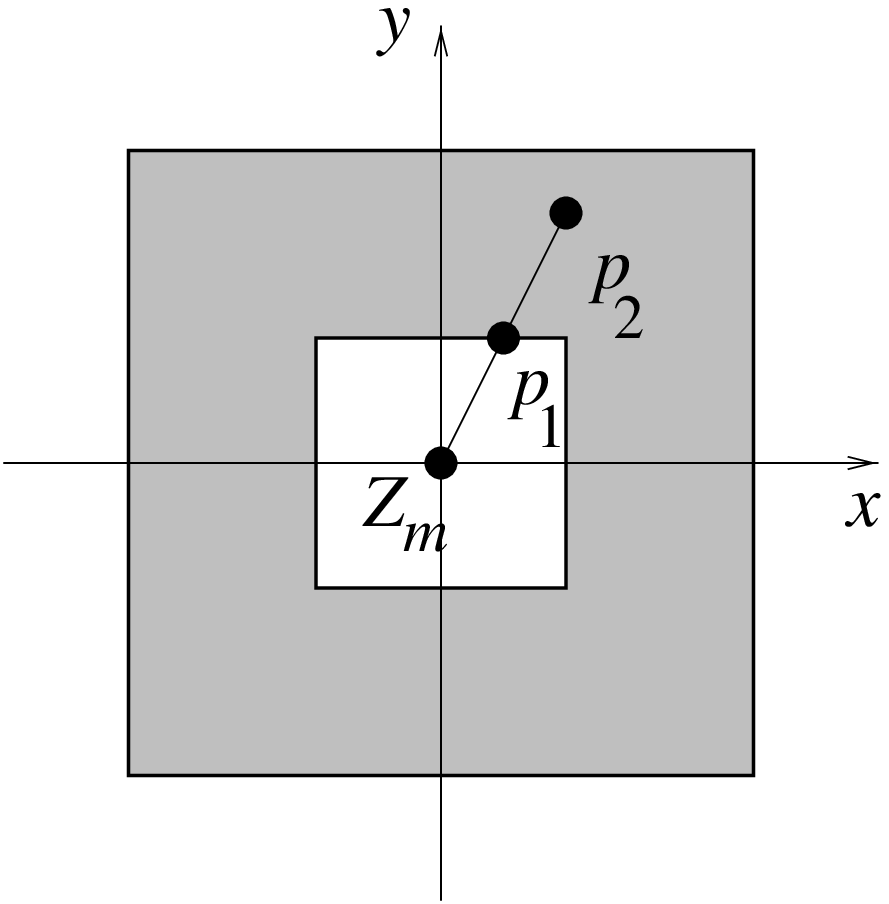,
        width=0.35\textwidth}}
    \caption{In the region $P$ (shown shaded), $p_2$ is dominated by $p_1$ and cannot be a local optimum.}
    \label{fig:widerspruch_rand}
  \end{center}
\end{figure}

In many cases, the following property of straight-line medians
can be applied for a reduction of the set of boundary segments that
we need to consider.
(See Figure~\ref{fig:widerspruch_rand} for an illustration.)
If $Z_m=(x_m,y_m)$ is the $L_1$ origin of $P$
and $p_1=(x_1,y_1)$ and $p_2=(x_2,y_2)$ are points in $P$,
we say that $p_1$
{\em dominates} $p_2$, if $p_1$ lies in the rectangle spanned
by $Z_m$ and $p_2$. 

\begin{lemma}
\label{le:dominate}
Let $p_1=(x_1,y_1)$ and $p_2=(x_2,y_2)$ be points in $P$. If 
$p_1$ dominates $p_2$, then $f(p_1)\leq f(p_2)$. 
\end{lemma}

\proof{
Suppose that $x_2>x_1\geq x_m$ and $y_2\geq y_1\geq y_m$.
Then $w(p_2)\geq w(p_1)\geq \frac{\mu}{2}$ and
$s(p_2)\geq s(p_1)\geq \frac{\mu}{2}$, so moving a center
from $p_2$ to $p_1$ cannot increase the objective value.
}

Using a plane-sweep algorithm, it is possible to identify the
non-dominated portions of the boundary in time $O(n \log n)$. If this
set has complexity $o(n)$, then we get a reduction of the overall
complexity.

\begin{figure}[htb]
  \begin{center}
    \leavevmode
    \centerline{\epsfig{file=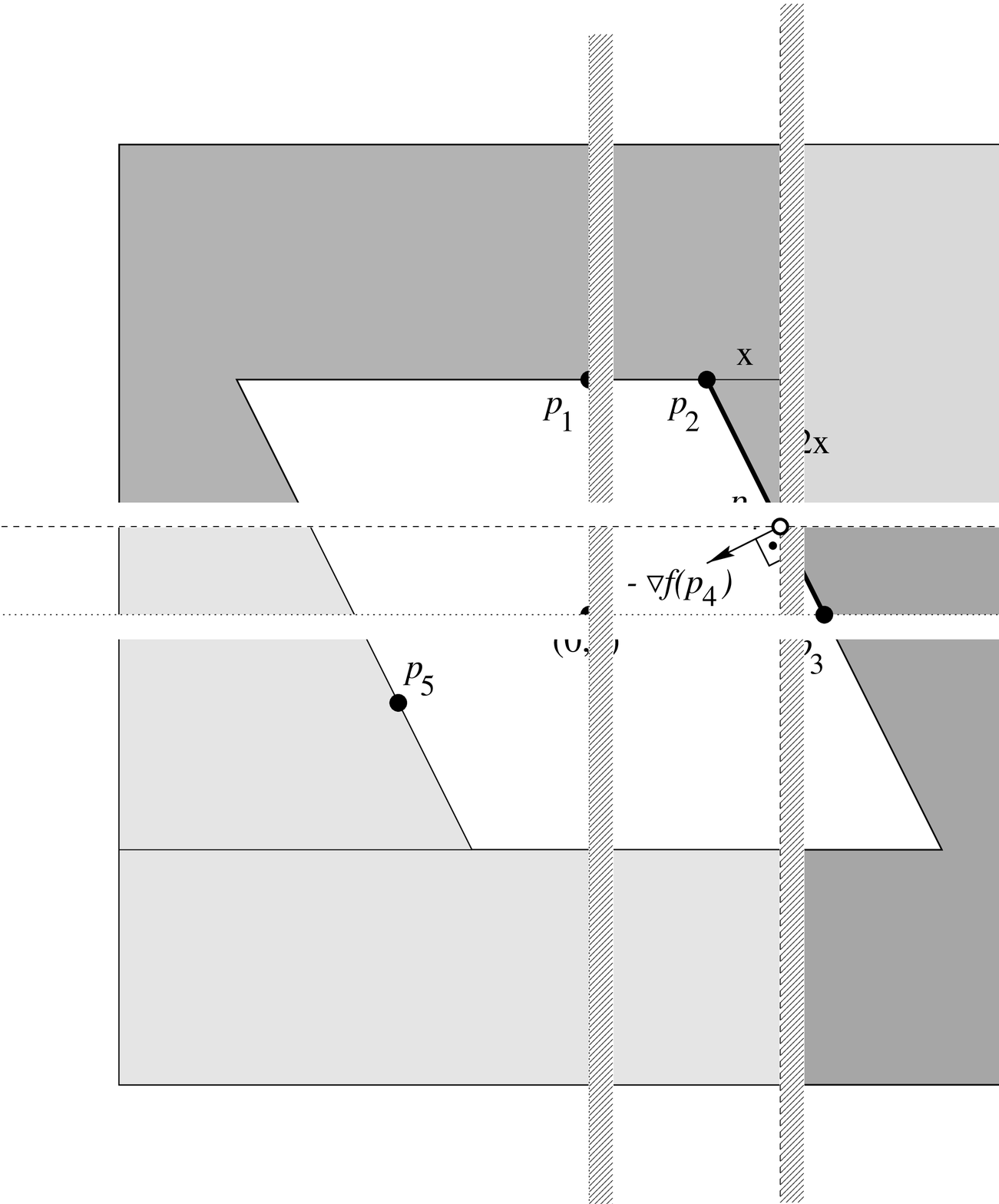, 
     width=0.6\textwidth}}
    \caption{The example from Figure~\protect{\ref{fig:beispiel}} is 
      analyzed for the case of straight-line $L_1$ distances.  There
      are two optimal center points: $p_4=(2 \sqrt{7} - \frac{9}{2},11
      - 4 \sqrt{7})$ and its mirror image, $p_5$.}
    \label{fig:beispiel_direkt}
  \end{center}
\end{figure}

A simple example of the problem solved in this section is shown in
Figure~\ref{fig:beispiel_direkt} (based on the example given in
Figure~\ref{fig:beispiel}). This example shows the necessity of
solving quadratic equations in computing the optimal solutions (points
$p_4$ and $p_5$).  The non-dominated points are $p_{1}$, $p_{2}$, and
$p_{3}$, as well as the points interior to the segment
$e_{2,3}=\overline{p_{2}p_{3}}$.

\old{Examining the edge $e_{2,3}$
yields a solution $p_4$ that is $x = 2 \sqrt{7} - 5$ 
right of $p_2$ and $2x$ down.
The objective values turn out to be
$f(p_{1}) = \frac{179}{72}$, $f(p_{2}) = \frac{182}{72}$, 
$f(p_{3}) = \frac{359}{144}$, and 
$f(p_4)=\frac{1362 - 448 \sqrt{7}}{72} = \frac{176,\dots}{72}$,
which is optimal.}

\section{Geodesic $L_1$ Distances in Simple Polygons}

In this section, we show how to compute in optimal ($O(n)$) time a
point that minimizes the average geodesic $L_1$ distance for a simple
polygon $P$ without holes.

\begin{figure}[hbt]
  \begin{center}
    \leavevmode
    \centerline{
    \epsfig{file=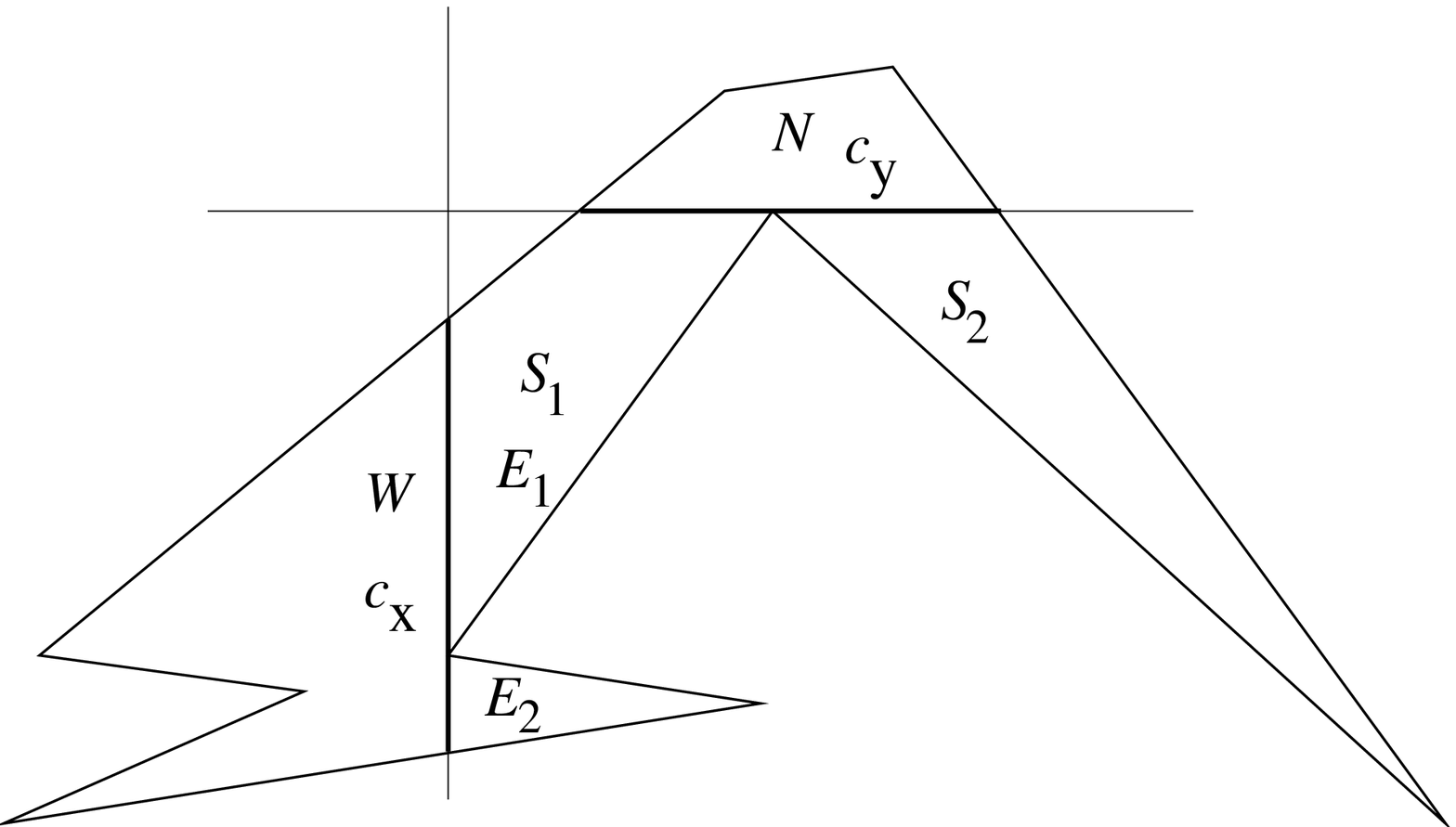, width=.6\textwidth}}
    \caption{Case 2: Both of the median chords are critical.}
    \label{fig:zwei_sprung}
  \end{center}
\end{figure}

{From} Lemma~\ref{le:geo}, we know that
the partial derivative $f_x(Z)=\frac{1}{\mu} (w(Z)-e(Z))$ 
vanishes if $w(Z)=e(Z)$. Therefore we 
consider the functions $w(Z)$ and $e(Z)$ that are well-defined even for 
points for which the gradient is not.
As $x$ increases, $w(Z)$ increases monotonically,
while $e(Z)$ decreases monotonically.
Note that $w(Z)$ may be discontinuous at critical vertices
of $P$: As shown in Figure~\ref{fig:zwei_sprung}, an entire
region may switch from ``east'' to ``west'' as a vertical chord
passes through a critical vertex.

However, even discontinuous behavior
at critical coordinates does not impair monotonicity of $w(Z)$ and $e(Z)$,
so there is still a well-defined {\em vertical median chord} $c_x$ at 
some $x$-coordinate $x_m$ such that $w(Z_1)<e(Z_1)$ 
for all
$Z_1=(x_1,y_1)$ with $x_1<x_m$ (implying $f_x(Z_1)<0$ just left of $x_m$), and
$w(Z_1)>e(Z_1)$ 
for all $Z_2=(x_2,y_2)$ with $x_2>x_m$ (implying $f_x(Z_2)>0$ just right
of $x_m$).
Similarly, there is a unique {\em horizontal median chord} $c_y$
at $y$-coordinate $y_m$.  Again we call $Z_m=(x_m,y_m)$ the 
$L_1$ {\em origin} of~$P$.

In the following, we use the structure of simple polygons to show that
the locally optimal point $Z_m$ has to belong to the feasible
region~$P$ (possibly on the boundary of~$P$), implying that it is a
unique global optimum.

\begin{theorem}
\label{th:feasible}
The point $Z_m$ is feasible (lies in $P$) and thus a unique global optimum,
minimizing the average $L_1$ geodesic distance to points in~$P$.
\end{theorem}

\proof{ The chord $c_x$ subdivides $P$ into two pieces: let $E$ denote
  the part to the right (``east'') of $c_x$, and $W$ the part to the
  left (``west'') of $c_x$. Note that $E$ or $W$ may consist of two or
  more connected components only if $c_x$ passes through a critical
  vertex.
  Similarly, the chord $c_y$ subdivides $P$ into the region $N$
  (``north'') that lies above $c_y$, and the region $S$ (``south'')
  that lies below~$c_y$.  
  
  We claim that $c_x$ and $c_y$ intersect at a point ($Z_m$) inside
  $P$.  The proof is by contradiction; assume that $Z_m$ lies outside
  $P$.  We will distinguish the following cases.

{\bf Case 0:
Neither $c_x$ nor $c_y$ are critical.} If $Z_m\not\in P$,
then simplicity of $P$ implies that
the two chords subdivide $P$ into three pieces; 
this means that precisely one of the pieces
$W$ and $E$ has nonempty intersection with one of the pieces 
$N$ and $S$.  
Without loss of generality, assume that the two chords subdivide $P$
into the three pieces, $E$, $N\cap W$, and $S$, as shown in
Figure~\ref{fig:kein_sprung}.  Since $P$ is a nondegenerate polygon,
the corresponding areas, $\mu(E)$, $\mu(N\cap W)$, and $\mu(S)$, are
all positive, with $\mu(E)+\mu(S)+\mu(N\cap W)=\mu(P)=\mu$.  However,
the local optimality of $x_m$ implies that $\mu(E)=\frac{\mu}{2}$ and
the local optimality of $y_m$ implies that $\mu(S)=\frac{\mu}{2}$,
implying the contradiction that $\mu(N\cap W)=0$.

\begin{figure}[hbt]
  \begin{center}
    \leavevmode
    \centerline{
    \epsfig{file=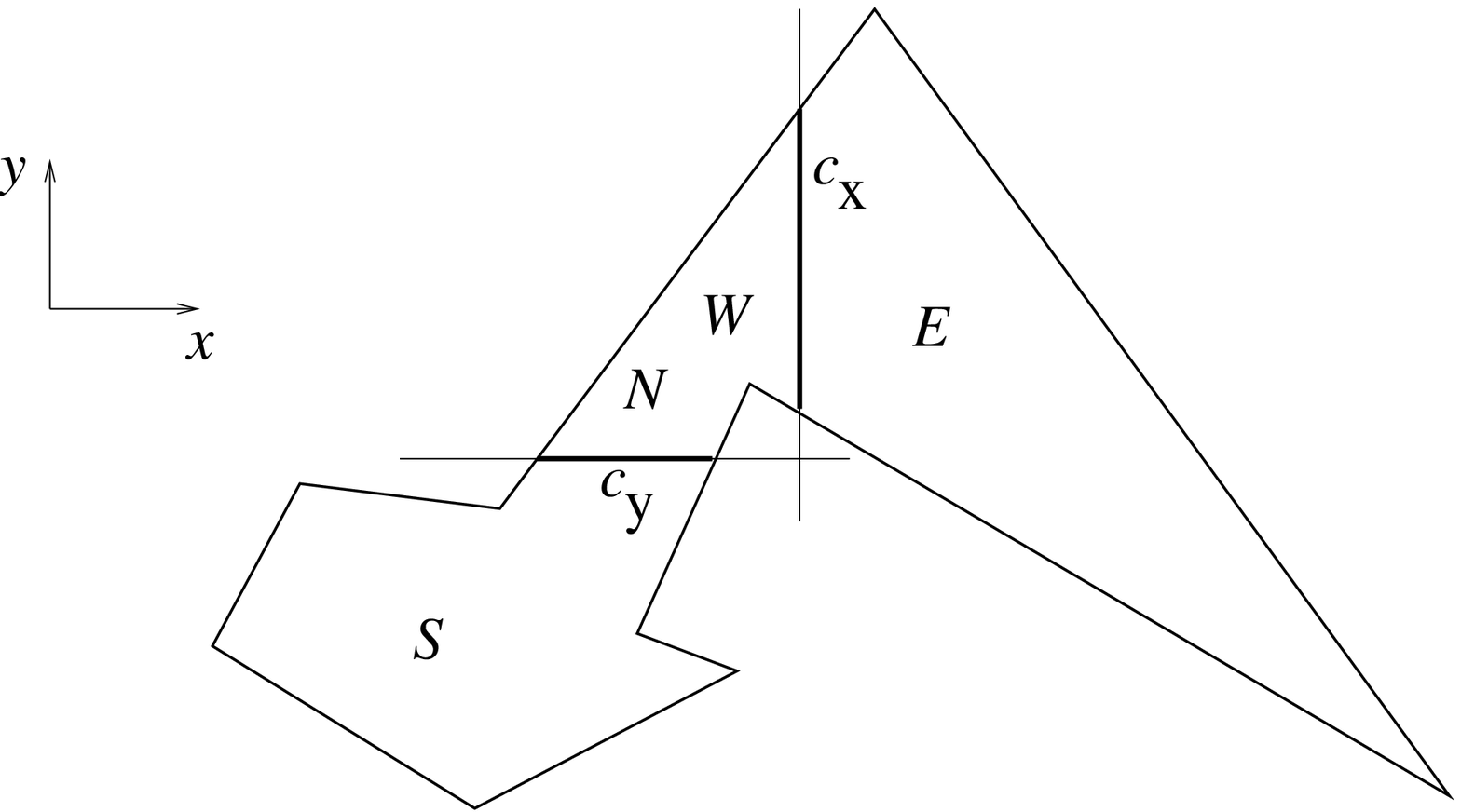, width=.6\textwidth}}
    \caption{Case 0: Neither of the median chords $c_x$ and $c_y$
      is critical.}
    \label{fig:kein_sprung}
  \end{center}
\end{figure}
  
{\bf Case 1:
Exactly one of $c_x$ and $c_y$ is critical.} 
Without loss of generality,
assume that $c_y$ passes through a critical vertex of $P$
that is a local maximum of the boundary of $P$, as in
Figure~\ref{fig:ein_sprung}.
As in Case~0, the (noncritical) vertical chord $c_x$ partitions $P$
into two pieces, $E$ and $W$, each of area $\frac{\mu}{2}$. Also,
$c_y$ partitions $P$ into one ``upper'' piece $N$ and two ``lower''
pieces, $S_1$ and $S_2$, each with positive area.  (In degenerate
situations, $c_y$ may pass through multiple critical vertices,
resulting in multiple lower pieces; our arguments includes
this case by considering additional pieces as part of $S_2$.)
The assumption that $c_x$ and $c_y$ do not
cross inside $P$ implies that either $W$ or $E$ is a strict subset of
$S_1$, $S_2$, or $N$.  This is a contradiction, since
$\mu(W)=\mu(E)=\frac{\mu}{2}$, and $\mu(N)=\frac{\mu}{2}$,
$\mu(S_1)<\frac{\mu}{2}$, and $\mu(S_2)<\frac{\mu}{2}$.

\begin{figure}[hbt]
  \begin{center}
    \leavevmode
    \centerline{
    \epsfig{file=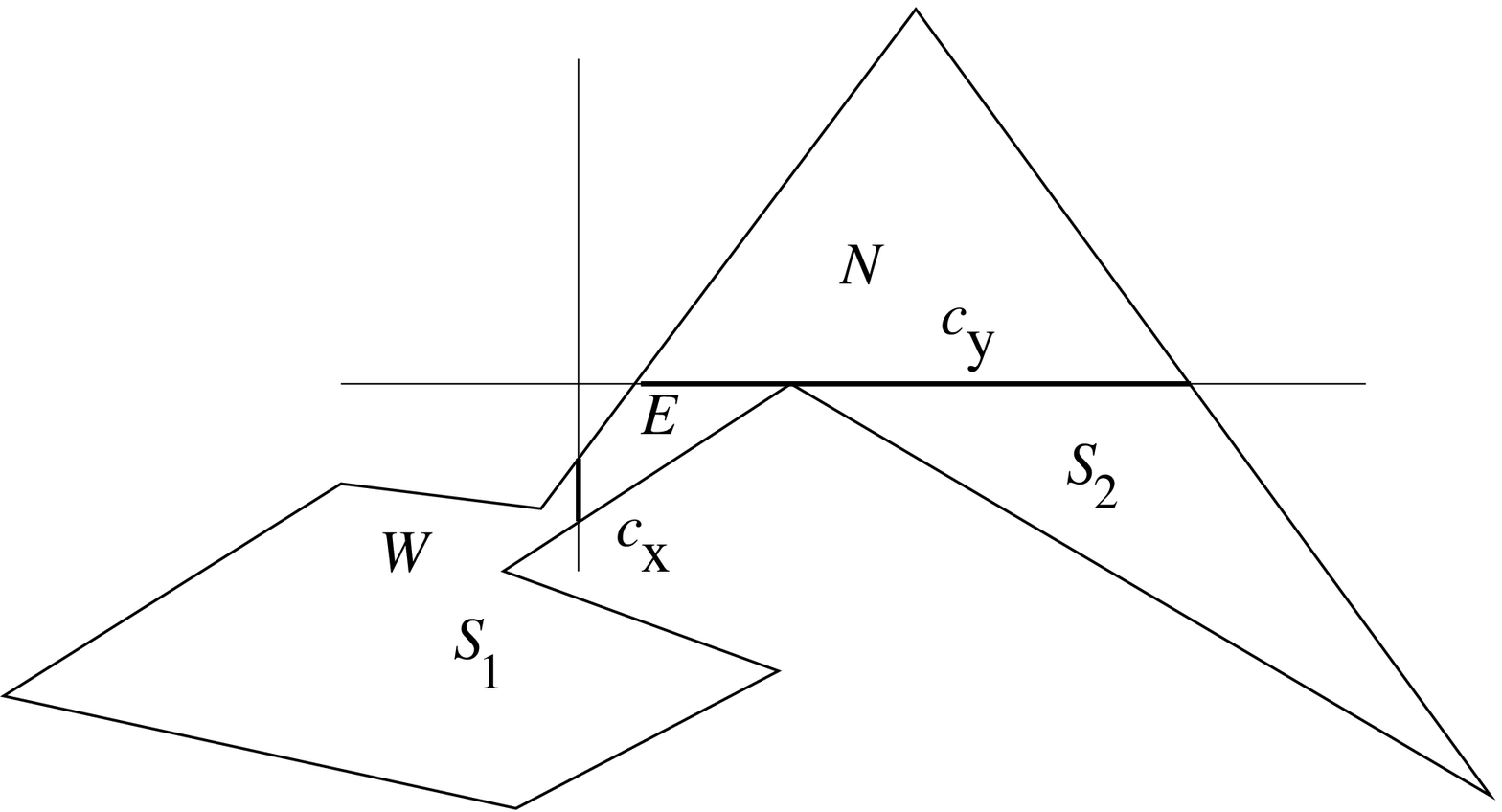, width=.6\textwidth}}
    \caption{Case 1: Exactly one of the median chords is critical.}
    \label{fig:ein_sprung}
  \end{center}
\end{figure}

{\bf Case 2: Both $c_x$ and $c_y$ are critical.}  Without loss of
generality, assume that $c_x$ passes through a critical vertex of $P$
that is a local minimum of the boundary, while $c_y$ passes through a
critical vertex that is a local maximum of the boundary, as shown in
Figure~\ref{fig:zwei_sprung}.  As in Case~1, the horizontal chord
$c_y$ subdivides $P$ into a ``northern'' piece $N$, and two
``southern'' pieces, $S_1$ and $S_2$, each with positive area.
Similarly, the vertical chord $c_x$ subdivides $P$ into a ``western''
piece $W$, and two ``eastern'' pieces, $E_1$ and $E_1$, each with
positive area.  We assume further, without loss of generality, that
the critical vertex through which $c_x$ passes lies within the piece
$S_1$, as shown in the figure.  We have a contradiction in the fact
that $\mu(N)=\mu(W)=\frac{\mu}{2}$, while the set $P\setminus (N\cup
W)$ has positive area greater than $\mu(S_2)$.}

\begin{theorem}
\label{th:geo.linear}
The point $Z_m$ can be computed in linear time.
\end{theorem}

\proof{
We describe how to compute the $x$-coordinate $x_m$ of $Z_m$; the 
$y$-coordinate is found in a similar manner.

In linear time (using Chazelle's algorithm \cite{chaz}), we build
the vertical trapezoidization of $P$, which is defined by drawing
vertical chords through every vertex of $P$.
Each piece, $\tau_i$, of the resulting subdivision
is either a vertical-walled
trapezoid or a triangle having one side vertical (such a triangle
can be considered to be a degenerate vertical-walled trapezoid).
Consider the adjacency graph ${\cal G}$ of these pieces $\tau_i$ (i.e., 
the planar dual of the trapezoidization);
because $P$ is a simple polygon, ${\cal G}$ is a tree. 


\begin{figure}[htb]
 \begin{center}
  \leavevmode
  \centerline{\epsfig{file=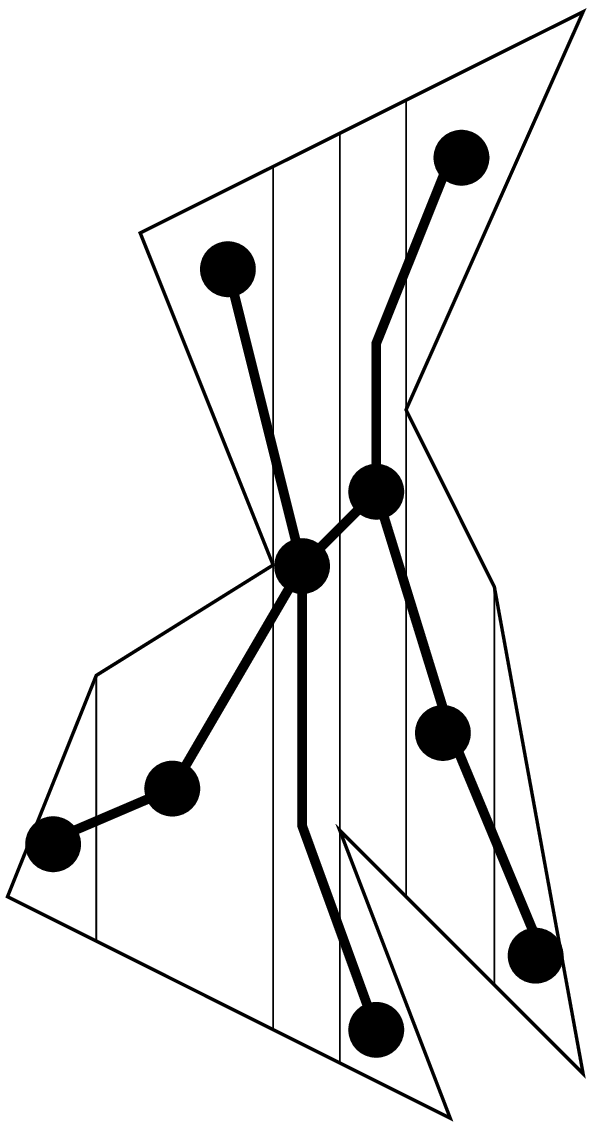, 
  width=0.30\textwidth}}
  \caption{A median trapezoid corresponds to a median in a node-weighted tree,
with node weight representing trapezoid area.}
  \label{fig:median}
 \end{center}
\end{figure}

Let $\tau_m$ denote the trapezoid containing the vertical chord $c_x$
through $Z_m$.  (We assume, without loss of generality, that $c_x$ is
not one of the vertical walls of $\tau_m$; the degenerate case is
readily handled by similar arguments.)  Let $C_{\max}$ be a connected
component of $P\setminus \tau_m$ that has maximum area; let
$\tau_{\max}$ be the unique trapezoid within $C_{\max}$ that is
(vertical wall) adjacent to $\tau_m$.  The area of $C_{max}$ cannot
exceed $\mu/2$, by the local optimality criterion.  (Moving $Z_m$ from
$\tau_m$ by an infinitesimal $\varepsilon$ into $\tau_{\max}$ would
reduce the distance to $Z_m$ by $\varepsilon$ for a set of points of a
total area more than $\mu/2$, while increasing it by at most
$\varepsilon$ for a set of points of total area less than $\mu/2$.)
Thus, $\tau_m$ corresponds to what is called a {\em median} node in
the weighted tree~${\cal G}$, whose nodes are weighted by the areas of
the corresponding trapezoids. See Figure~\ref{fig:median}.

A median in a weighted tree can be computed in linear time (e.g., see
Goldman~\cite{Gold71}; the oldest reference appears to be from
Hua~\cite{h-ammwh-62}). This allows us to compute in linear time a
trapezoid $\tau_m$ that contains the vertical chord~$c_m$.

Once $\tau_m$ has been identified, it is easy to compute $x_m$ (the
$x$-coordinate of $c_x$).  We desire the solution to the equation
$\mu_W + q(x_m)=\mu_E + (\mu(\tau_m)-q(x_m))$, where $\mu_W$ (resp.,
$\mu_E$) is the area of all components of $P\setminus \tau_m$ that are
adjacent to the left (resp., right) wall of $\tau_m$, $\mu(\tau_m)$ is
the area of $\tau_m$, and $q(x_m)$ is the area of the portion of
$\tau_m$ to the left of coordinate $x_m$.  It is easy to see that $q(\cdot)$
is a quadratic function; thus, $x_m$ is readily computed as a root
of a quadratic equation. Since $\mu_W$ and $\mu_E$ are readily computed
in linear time once $\tau_m$ is identified, the computation of $x_m$
takes linear time in total.  Similarly, we compute $y_m$ in linear time.}

\section{Geodesic $L_1$ Distances in Polygons with Holes}

Now we discuss an even more complicated case, which
arises when considering geodesic $L_1$ distances
in polygonal regions $P$ that may have holes.
Again, we analyze the set of locally optimal points:
as long as a potential center can be moved in some
axis-parallel fashion that lowers the average $L_1$ geodesic distance
to all the points, it cannot be optimal.

The local optimality of a point $Z$
is closely related to the subdivisions that it induces:
for local optimality in the $x$-direction, 
the subdivision into $W(Z)$ and $E(Z)$ needs to be area-balanced; for 
local optimality in the $y$-direction, 
the subdivision into $N(Z)$ and $S(Z)$ needs to be area-balanced. 
(Refer to Lemma~\ref{le:geo}.)
The boundary between $W(Z)$ and $E(Z)$ is formed by bisectors
in the shortest path map, SPM($Z$) with respect to $Z$. It follows from basic properties of shortest
path maps that the total complexity of this boundary is $O(n)$.
(See, e.g., \cite{m-naspo-91}.)

As we showed in Lemma~\ref{le:cubic}, there is a neighborhood
for each point $Z\in P$ in which the objective function $f$
is cubic, provided that no bisector for $Z$ meets a boundary
vertex. This motivates the following lemma:

\begin{lemma}
\label{le:equiv_map}
There is a subdivision of $P$ of worst-case complexity $I=\Theta(n^4)$,
such that $f$ is a cubic function within each face of the subdivision.
\end{lemma}

\proof{
Lemma~\ref{le:cubic} implies that we are done if we can
compute a subdivision of the claimed complexity
such that we can move continuously between
any two points in the interior of a connected cell of the subdivision,
without any bisector encountering a vertex of the polygon 
during this motion.
Provided that there is a position $Z$ for which a bisector encounters
a vertex $v$ of the polygon, this
vertex $v$ is contained in $W(Z)$ as well as in $E(Z)$.
Thus, there are two topologically
different shortest paths from $Z$ to $v$, one fully contained in
$W(Z)$, the other contained in $E(Z)$. This implies that there
are two topologically different shortest paths from $v$ to $Z$, i.\,e.,
$Z$ must lie on a watershed bisector in SPM($v$). Therefore, the required subdivision
is obtained by considering the $O(n)$ watershed bisectors in each of the $O(n)$
shortest path maps with respect to polygon vertices. Each shortest path map has a complexity of
$O(n)$, so the subdivision is defined by the overlay
of $O(n^2)$ line segments, yielding an arrangement of worst-case 
complexity $I=O(n^4)$.  The example in Figure~\ref{fig:n4cells}
shows that even in the case of simple polygons,
this bound on $I$ is tight in the worst case.  (Chiang and
Mitchell~\cite{cm-tpesp-99} have studied similar arrangements that
arise in overlaying shortest-path maps in the Euclidean shortest-path
metric.)  

\begin{figure}[htb]
  \begin{center}
    \leavevmode
    \centerline{\epsfig{file=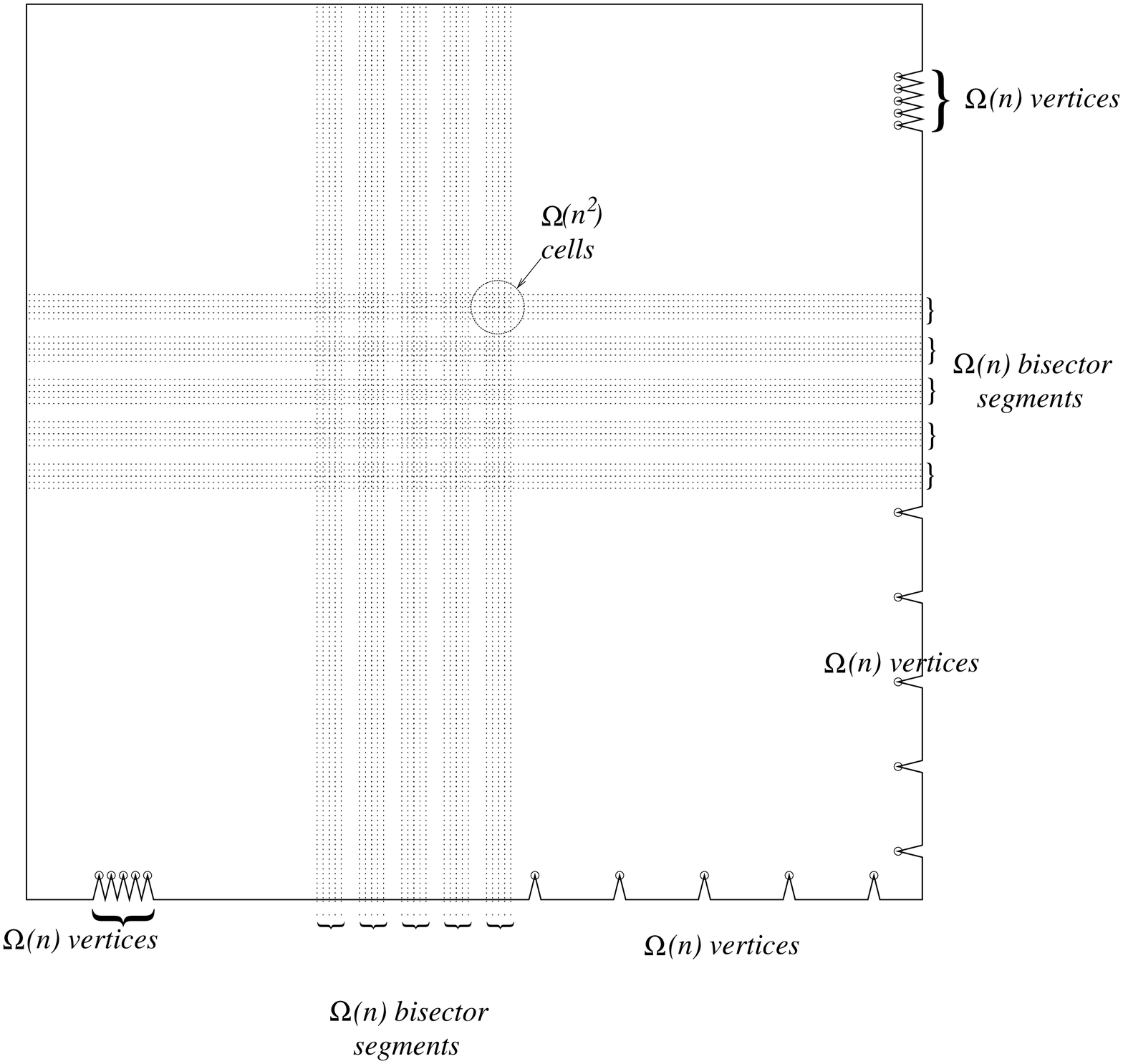,
        width=0.65\textwidth}}
    \caption{An example in which the overlay subdivision has complexity $I=\Theta(n^4)$.}
    \label{fig:n4cells}
  \end{center}
\end{figure}
}

Considering the local optima for each cell
of the arrangement allows us to obtain the following:

\begin{theorem}
\label{th:geo.holes}
For geodesic $L_1$ distances, a feasible point $Z^*=(x^*,y^*)$
in a polygonal region $P$ with holes that minimizes the average 
(geodesic $L_1$) distance $f$ to all points in $P$ can be found in
worst-case time $O(I+n \log n)$.
\end{theorem}

\proof{ The search for optimal solutions proceeds on a cell by cell
  basis, for each of the $O(I)$ cells in the overlay arrangement.  The
  overlay arrangement can be computed in time $O(I+n\log n)$, using
  known algorithms (\cite{b-oafsi-95,ce-oails-92}).  (We utilize the
  perturbation argument given in Section~\ref{sec:prelim} in order to
  be able to assume, without loss of generality, that the bisectors
  are all polygonal curves, not regions of nonzero area.)  For each
  cell, we spend constant time conputing the $O(1)$ candidate (local)
  optima inside and on the boundary of the cell. The function
  parameters for $f$ within each cell can be determined in total time
  $O(I+n \log n)$ by traversing the arrangement (e.g., by depth-first
  search in the planar dual graph of the arrangement) and doing simple
  $O(1)$-time updates when changing from one cell to a neighboring
  cell. After determining the $O(I)$ candidate locations, we can
  determine a best among them by computing their objective values,
  again in total time $O(I+n \log n)$, by performing incremental
  updates to the objective function values during a traversal of the
  arrangement. For any given cell of the arrangement, if there is a
  local minimum interior to the cell, the gradient $\grad f$ must
  vanish. Because $f$ is the sum of two cubic functions,
  $f_1(x)$ and $f_2(y)$, within the cell, this means that we get
  a system of two quadratic equations (both components of the gradient
  must be zero) with two variables ($x$ and $y$). Such a system can be
  solved in constant time using radicals.
  
  Similarly, we can determine the local optima with respect to
  variation along a boundary segment of a cell.  For each segment, the
  gradient needs to be orthogonal to the segment. As in the
  straight-line case, this yields a quadratic equation that can be
  solved in constant time.
  
  Finally, there are $O(I)$ vertices in the arrangement, each of which
  we consider to be candidates. 
 
  In total, then, we have examined $O(I)$ candidate local minima, in
  time $O(I+n\log n)$.}

\section{Multiple Centers}
\label{sec:many-centers}

We now discuss the $k$-median problem of placing $k$ centers into a
polygonal region $P$, such that the overall average distance of all
points $p\in P$ to their respective closest centers is minimized.  
We consider $k$ to be part of the input and potentially large.

\begin{theorem}
\label{th:NPc}
For polygons $P$ with holes,
it is NP-hard to determine a set of $N$ centers that
minimizes the average geodesic $L_1$ distance from the points in $P$
to the nearest center.
\end{theorem}

\proof{
Our construction
uses a reduction from {\sc Planar 3Sat}, which was shown to 
be NP-complete by Lichtenstein~\cite{lich}.
We recall that a 3SAT instance $I$
is said to be an instance of {\sc Planar 3SAT},
if the following bipartite graph $G_I$ is planar:
each variable $x_i$ and each clause $c_j$ in $I$ is represented
by a vertex in $G_I$; two vertices are connected if and only
if one of them represents a variable that appears in the
clause that is represented by the other vertex.
See Figure~\ref{fig:planar-3-sat}. The variable-clause incidence graph
can be embedded in the plane without any crossing edges.

First, the planar graph $G_I$ corresponding to an instance $I$ of {\sc
Planar 3Sat} with $n$ variables and $m=O(n)$ clauses
is represented in the plane as a {\em planar rectilinear
layout}, with each vertex corresponding to a horizontal line segment,
and each edge corresponding to a vertical line segment that
intersects precisely the line segments corresponding to the two
incident vertices. There are well-known algorithms (e.g.,
\cite{rose_tar}) that can achieve such a layout in $O(n)$ time and
$O(n)$ space.  See Figure~\ref{fig:planar-3-sat}.
We assume that the eventual layout is scaled appropriately by
a factor of size $\Theta(n)$,
such that the overall size is $\Theta(n^2)$.

\begin{figure}[htb]
 \begin{center}
  \leavevmode
  \hspace{-40pt}
  \centerline{\epsfig{file=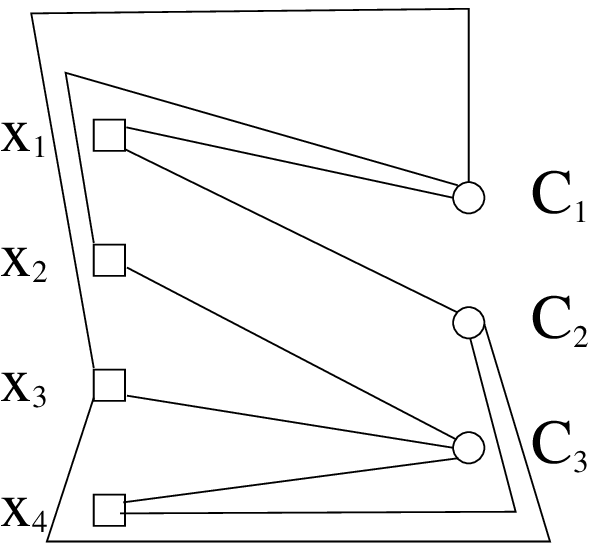, width=.3\textwidth}}
    \centerline{
    \epsfig{file=figures/empty.eps, height=.2in}}
  \centerline{\epsfig{file=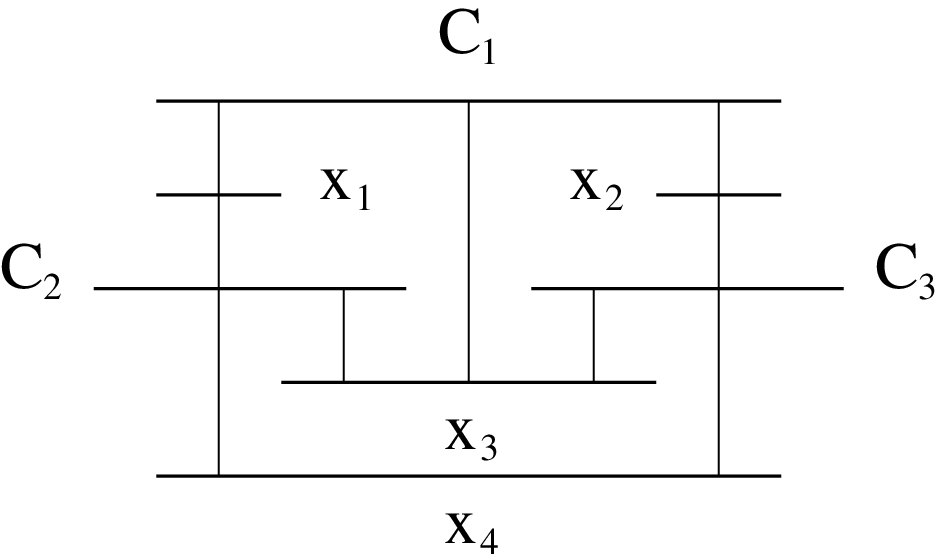, width=0.5\textwidth}}
  \caption{The graph $G_I$ for the Planar 3SAT instance
  $I=(x_{1} \vee x_{2} \vee \bar x_{3}) \wedge (\bar
    x_{1} \vee x_{3} \vee x_{4}) \wedge (\bar x_{2} \vee \bar x_{3} \vee
   \bar x_{4})$, and its geometric representation.}
  \label{fig:planar-3-sat}
 \end{center}
\end{figure}

Next, the layout is modified such that the line segments corresponding
to a vertex and all edges incident to it are replaced by a loop -- see
Figure~\ref{fig:einbettung-schleife} (top).  At each vertex
corresponding to a clause, three of these loops (corresponding to the
respective literals) meet.  Finally, the edges of any loop $i$ are
replaced by a sequence of $3c_i$ small squares 
(say, of size $\varepsilon=O(1/n)$)
that are spaced apart at a constant distance (say, $d=O(1)$) and are interconnected by
narrow corridors (say, of width $\varepsilon^{11}$) that have small
enough total area that they do not greatly influence the overall
average distance: The total area of all corridors is
$O(n^3*\varepsilon^{11})=O(\varepsilon^{8})$,
and the maximum distance between two points in corridors
is $O(n^3)$, so the integral of pairwise distances over all corridor
points is $O(\varepsilon^{5})$.
Along each variable loop, the sequence of $3c_i$ squares
is labeled ``false'' (index 0 mod 3 in the sequence),
``true'' (index 1 mod 3 in the sequence), and ``nil''
(index 2 mod 3 in the sequence), in succession.

Similarly, each vertex for a clause is replaced by a
single small square and linked to the adjacent variable loops by three
narrow corridors of length $d$, with adjacency encoding
to the corresponding literal, i.e., connecting the clause square
to a ``true'' square for an unnegated and to a ``false''
square for a negated literal. Note that  no clause square is
adjacent to a ``nil'' square.
See Figure~\ref{fig:einbettung-schleife} (bottom) for the overall picture. 

\begin{figure}[htb]
 \begin{center}
  \leavevmode
  \hspace{-20pt}
  \centerline{
  \epsfig{file=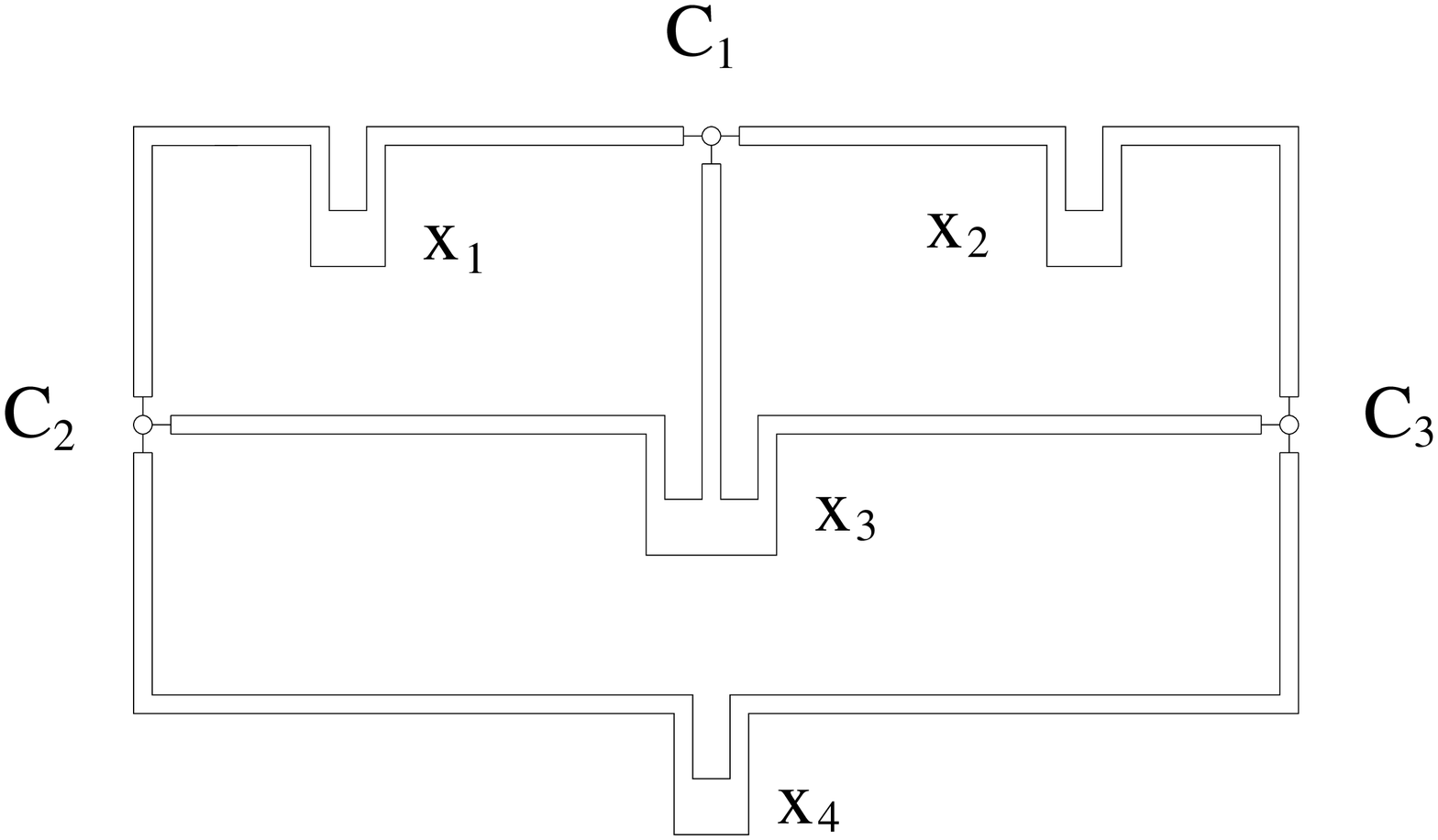, width=.4\textwidth}}
    \centerline{
    \epsfig{file=figures/empty.eps, height=.2in}}
  \centerline{
  \epsfig{file=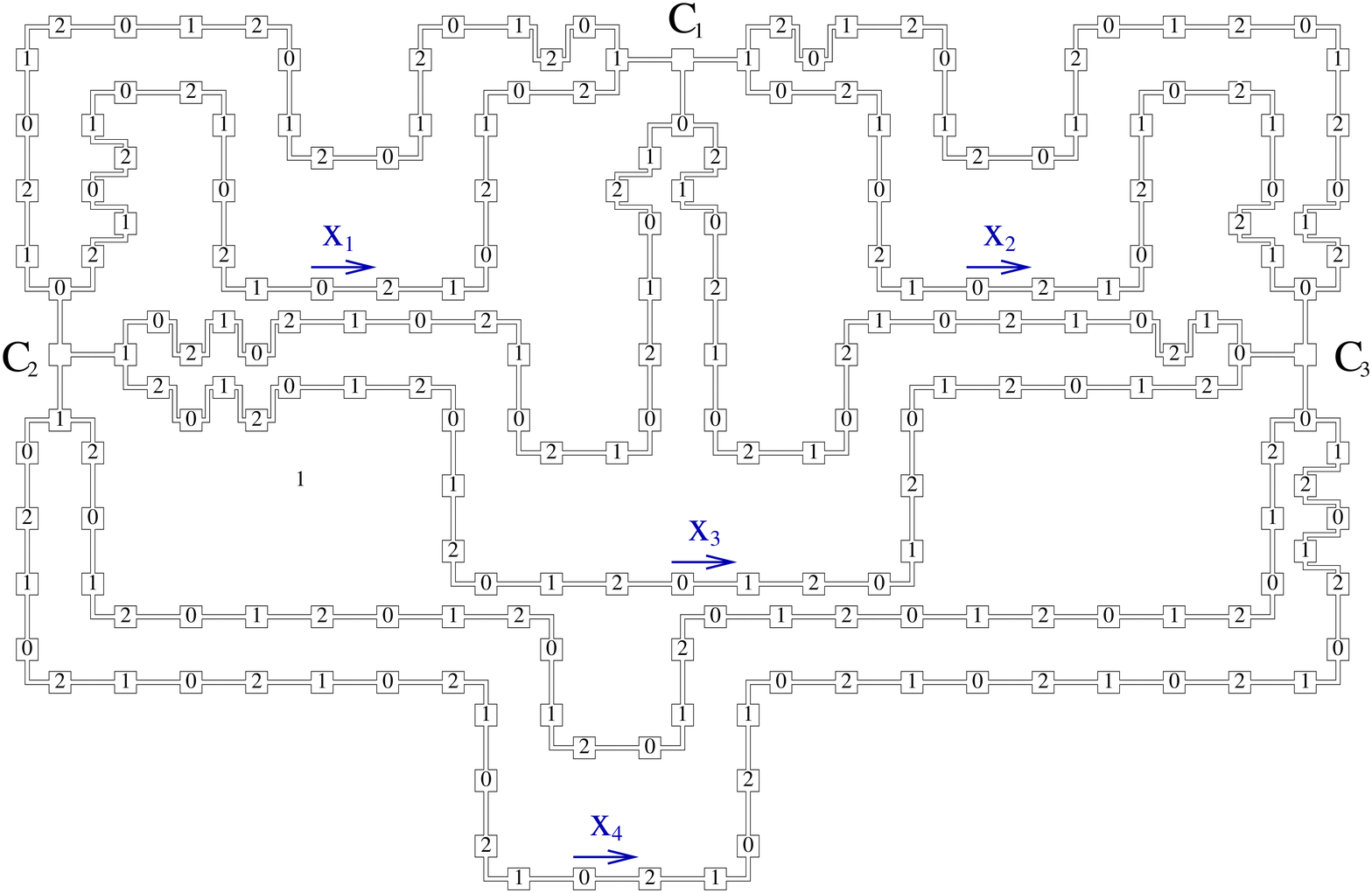, width=.75\textwidth}}
  \caption{Replacing variables by loops (top); final polygon (bottom).
   Numbers 0, 1, 2 indicate the ``true'', ``false'' and ``nil'' squares.}
  \label{fig:einbettung-schleife}
 \end{center}
\end{figure}

Let $N=3c=O(n^2)$ be 
the total number of squares in all variable loops, and consider
the placement of $N$ centers. Because the total area of the corridors is
small enough, neglecting them in the following discussion changes
the resulting overall average distances by not more than $O(\varepsilon^5)$.
Furthermore, we assume without loss of generality that the placement of
centers is locally optimal, so any center is placed as a median
of the squares closest to it. It is readily checked (making use of the
constant distance between adjacent squares and the negligible area
of corridors) that this
allows us to assume that all centers have been placed inside of
squares.

Now it is easy to estimate the overall average distance:
We get an average distance of 
$$D:=\frac{1}{3c}\left(\Theta(\varepsilon^4)+\sum_{s=1}^{3c}n_sd\right),$$
where $n_sd$ is the distance of the midpoint of square $s$ to the midpoint
of the closest square containing a center point; by construction,
each $n_s$ is a nonnegative integer. As there can be at most
$c$ squares with $n_s=0$, we conclude that $\sum_{s=1}^{3c}n_s\geq 2c$,
with equality if and only if each square is either occupied by
or next to a square with a center. For this reason, we call a 
square $s$ {\em covered}, if and only if $n_s\leq 1$. It follows from the above
description that there is a distribution of centers with an
average distance of $\frac{1}{3c}\left(\Theta(\varepsilon^4)+2c\right)$, 
if there is a covering positioning;
on the other hand, we see that the average distance must be at least
$\frac{2c+1}{3c}$ if there is no covering positioning of centers.

To establish the claim of NP-hardness, we show in the following
that there is a satisfying truth assignment of the instance $I$, if and only if
there is a distribution of centers to squares, such that all squares
are covered. 

First assume that there is a covering.
Consider the set of $c_i$ ``nil'' squares
in a variable loop. Clearly, no two of them 
can be covered by the same center; by construction, no ``nil'' square
can be covered by a center in a clause square. This implies
that each variable $i$ requires precisely $c_i$ centers to be
placed in its squares, so all $c$ centers
must be placed on variable squares. 
As all $3c_i$ squares in loop $i$ must be covered
by its $c_i$ centers, and no center can cover more than three
of the squares, we conclude that each center covers precisely three
of the variable squares. Thus, all centers in a variable loop must
be uniformally chosen to be all ``true'', all ``false'' or all ``nil''.
Finally, each clause square must be covered from one of its
adjacent variable squares, which means that setting
variable $i$ to the value indicated by the respective truth value
satisfies the clause. Thus, all clause squares can only be satisfied
if there is an overall satisfying truth assignment $I$.

Conversely, it is clear that for a satisfying truth
assignment for instance $I$, placing $c_i$ centers in the $c_i$
``true'' or ``false'' squares of variable $i$ (corresponding to
the truth setting of variable $i$) yields a covering distribution
of centers.

This concludes the proof.}

Note that the above proof can also be applied to the case of
geodesic $L_2$ distances, or when minimizing the maximum
distance instead of the average distance; furthermore, the underlying proof
technique can also be applied to other types of location problems.
For example, see~\cite{fm-orvgr-03} for a game-theoretic scenario
in which two players try to claim as much area as possible
by placing centers, and the second player must place all of his points
after the first player has played all of her points.

\section{Conclusion}
\label{sec:conclusion}

In this paper, we have given the first exact algorithmic
results for the Fermat-Weber problem for a continuous set
of demand locations. We have shown that for $L_1$
distances in the plane, we can determine
an optimum center in polynomial time, with
the complexity ranging from $O(n)$ for the case
of geodesic $L_1$ distances in simple polygons,
to $O(n^2)$ for straight-line distances
in general polygonal regions, and $O(n^4)$
for geodesic $L_1$ distances in polygons with holes.
Our results rely on a careful understanding of the local
optimality criteria, together with the structure and combinatorics of
shortest path maps.

\paragraph{Extensions and Open Problems.}
\quad

(1) Our results can be extended to ``fixed orientation metrics'' defined
by any constant number of directions.  (The $L_1$ metric is the
special case in which the two fixed orientations are horizontal and
vertical.)  The local optimality conditions become more complex;
however, the inherent algebraic complexity remains the same, for any
metric whose disks are convex polygons.  This extension allows one to
approximate the Euclidean ($L_2$) case to any desired degree of
precision.

(2) Our local optimality conditions generalize to the case of
more general (non-uniform) nonnegative demand
densities $\delta(p)$ by using the following
observation. Regardless of the demand density function,
any center location $Z\in P$ induces a subdivision
of $P$ into $E(Z)$ and $W(Z)$, and into $N(Z)$ and $S(Z)$
by shortest-path bisectors. Then the local optimality condition
on $Z$ requires that $E(Z)$ and $W(Z)$, and $S(Z)$ and
$N(Z)$ are balanced in the following sense:
instead of requiring that 
$E(Z)$ and $W(Z)$, as well as $N(Z)$ and $S(Z)$, have the same 
area, the balance condition is that locally optimal points $Z$
must have the integrals $\int_{p\in W(Z)}\delta(p) dp$
and $\int_{p\in E(Z)}\delta(p) dp$,
and $\int_{p\in N(Z)}\delta(p) dp$
and $\int_{p\in S(Z)}\delta(p) dp$
the same. Points with these properties
are called {\em $\delta$-medians}. Similar
ideas can be used for describing boundary points.
If, for a particular $\delta$,
there is a limited number of $\delta$-medians,
they can be computed in polynomial time,
and it is possible to compare objective values
in polynomial time, then we can determine
a $\delta$-center for the given region.
This includes the case in which the demand function
is given by point weights in combination with a uniform demand
distribution over $P$, which is a problem formulated by
Wesolowsky and Love~\cite{wl-lfrda-71}.
It is also easy to see that the above methods can be applied
for the case in which $F\neq D$ and distances are straight-line $L_1$ distances.
Note that
geodesic distances are not well-defined in this case;
however, if we use a combination of straight-line distances outside of $F$
and geodesic distances inside of $F$, our methods still apply.

\begin{figure}[htb]
 \begin{center}
  \leavevmode
  \centerline{\epsfig{file=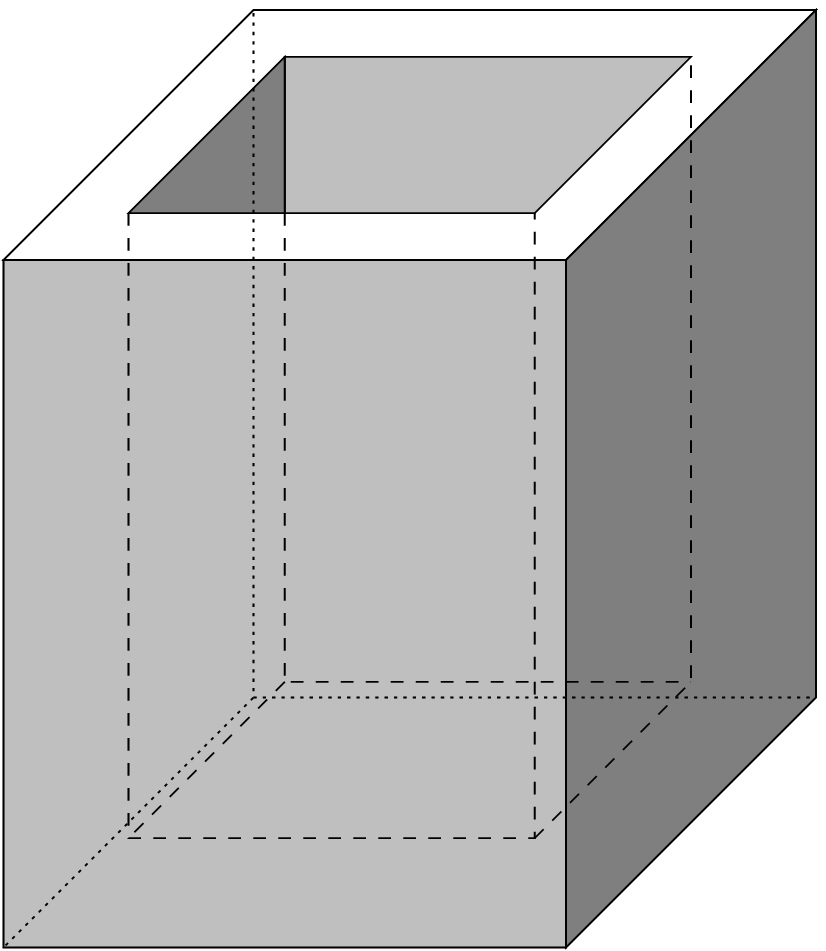, 
  width=0.2\textwidth}}
  \caption{In three-dimensional space, there may
    not be a feasible point that is a median of $P$ in all
    coordinates. In this example, the intersection of the three median
    planes (orthogonal to the $x$-, $y$-, and $z$-axes) is a point
    approximately at the center of the bounding box, not lying within
    the solid.}
  \label{fi:nosimple.3D}
 \end{center}
\end{figure}

(3) Our methods can also be applied in higher dimensions, by
generalizing the local optimality conditions and carrying through the
analysis in a very similar manner to the two-dimensional case.
Figure~\ref{fi:nosimple.3D} shows that a generalization of
Theorem~\ref{th:feasible}, however, does not hold in three-dimensional
space (since any axis-parallel plane cuts the region into not more
than two pieces), so we cannot use the same idea that allowed us in
the two-dimensional case to exploit simplicity in achieving a better
complexity than in the case of a polygon with holes.  However, we can
apply the technique of decomposing space into cells and studying the
objective function within each cell.  As in the two-dimensional case,
the objective function is cubic for each coordinate, if $P$ is a
polyhedral region.

(4) Our methods for searching for local optima should be extendable to
the case in which we have a constant ($k=O(1)$) number of centers,
e.g., $k=2$. The centers induce a subdivision of $P$ into several
``Voronoi regions'', corresponding to the set of points closest to
each center. Each center must be placed optimally with respect to its
region, which can be done by our methods.  Thus, we are done if we
have a suitable way to characterize the boundaries of Voronoi regions,
which consist of a number of bisectors.  While there are considerable
technical details to establish, we believe that this approach will
allow our results to generalize to multiple centers, and will lead to
an algorithm of complexity $O(n^{ck})$, for a small constant~$c$.

(5) It would be most interesting to discover an algorithm with
worst-case complexity better than $O(n^4)$ that can compute an
optimal center in the geodesic $L_1$ distance for polygons with holes.
Can we use geometric special structure to avoid examining all
potential local minima in the cells of the overlay arrangement?

\section*{Acknowledgments}
We would like to thank Arie Tamir, Horst Hamacher,
Justo Puerto,  Rainer Burkard, and Stefan Nickel for various
helpful comments and relevant references.
We also thank two anonymous referees for their many suggestions
that improved the presentation.
Finally, Iris Weber deserves our thanks for a heroic effort
in a critical situation.

\bibliographystyle{plain}
\bibliography{refs}  

\end{document}